\documentclass[12pt, runningheads]{llncs}
\usepackage{amssymb}
\usepackage{graphicx}
\usepackage{verbatim}
\usepackage{url}
\usepackage{enumerate}
\usepackage{comment}
\newcommand{\keywords}[1]{\par\addvspace\baselineskip
\noindent\keywordname\enspace\ignorespaces#1}

\usepackage{caption}
\usepackage[]{algorithm2e}

\usepackage{indentfirst}

\usepackage{listings}  
\usepackage{xcolor}

\usepackage{amssymb}
\usepackage{enumerate}

\usepackage{mathptmx}		
\usepackage{setspace}

\begin{document}
\mainmatter

\title{S$^3$CDM: A secret-sharing-scheme-based cyberattack detection model and its simulation implementation}

\author{Chi Sing Chum$\,^1$, Jia Lu$\,^1$, Claire Tang$\,^2$, and Xiaowen Zhang$\,^{1,*}$}


\institute{$^1$~Computer Science Dept., College of Staten Island, CUNY\\
2800 Victory Blvd, Staten Island, NY 10314, U.S.A. \\
$^2$~Hunter College High School\\
71 East 94th Street, New York, NY 10128, U.S.A. \\
$^*$ Corresponding Author Email: xiaowen.zhang@csi.cuny.edu 
}

\maketitle

\onehalfspacing 

\begin{abstract}

We design and develop a secret-sharing-scheme-based cyberattack detection model (S$^3$CDM) that can detect unauthorized or illegal activities (especially insider attacks) and protect sensitive information within complex network infrastructures of large organizations. The model splits a secret among a group of legitimate participants or components for authentication, integration and detection of unauthorized activities. Traditional Shamir's polynomial interpolation based and our own hash function based schemes are utilized in the model, they both are practical and efficient to make sure the communications between different components are secure and any unauthorized activities can be detected. The model offers a flexible multifactor authentication method to enhance the overall system security. Probability analysis \cite{CWZ23} shows that multiple component model is more resistant against cyberattacks than the single component one. To demonstrate the feasibility, we implement the S$^3$CDM in three parts on Google Cloud Platform, i.e., the frontend UI (User Interface) running on an HTTP server, the backend individual services written in Python, and a PostgreSQL database. Docker is used to manage the start and stop of individual services and their URLs. We demonstrate how to use the UI with a use case of simulation of broken path in details.



\end{abstract}

\section{Introduction}

A cyber threat \cite{JBWSS16} is ``any circumstance or event with the potential to adversely impact organizational operations (including mission, functions, image, or reputation), organizational assets, individuals, other organizations, or the Nation through an information system via unauthorized access, destruction, disclosure, or modification of information, and/or denial of service.'' Even worse, the threat could come from an insider that is supposed to be a trusted entity in an organization, with access to sensitive information and critical resources. \cite{HTGEO19,SD16} classify insider threats and detection approaches into several classes. Insider threats are one of the most challenging attacks. Secret sharing schemes are a much better solution to defeat or mitigate insider threats. 

Living in the cyber realm, almost everything is handled and stored in networked computers. All of our information, from everyday routine  activities, such as shopping, banking, and health care, to very important critical infrastructure runnings, such as electric grid, water system, transportation system, and homeland security, are being stored in the cloud-based remote servers. Our national critical infrastructures, manufactures, homeland security, financial institutions, and countless other industries are all controlled by computer systems. Hackers or attackers are persistently using any means, including social engineering, attacking networks, penetrating authentication systems of sensitive information and critical actions. They utilize bots, launch denial of service attacks, and send out millions of phishing messages/emails. Furthermore, in a lot of cases, the attackers are the insiders of organizations. It is too risky to let one individual or one entity hold the complete control or key to an access of critical computer systems. A secret sharing scheme is a mechanism to safeguard a secret with the added feasibility of availability and access control. There are a lot of applications of secret sharing schemes. Intrusion detection and threat detection systems are an extremely important part of cybersecurity measure. 

We design and develop a cyberattack detection model, which is based on secret sharing schemes, for detecting unauthorized or illegal activities and safeguarding sensitive information within complex network infrastructures of large organizations. The model splits a secret among a group of participants or components, uses secret sharing schemes for authentication, integration and detection of unauthorized activities. Probability analysis \cite{CWZ23} shows that multiple component model is more resistant against cyberattacks than the single component one. Traditional polynomial interpolation-based and hash function based schemes are utilized in the model, they both are practical and efficient to make sure the communications between different components are secure and any unauthorized activities can be detected. To demonstrate the feasibility, we implement the model in three parts on Google Cloud Platform, i.e., the frontend UI (User Interface) running on an HTTP server, the backend individual services written in Python, and a PostgreSQL database. We also demonstrate how to use the UI with a use case.

The rest of paper is organized as follows. In Section \ref{SSS}, we explain secret sharing schemes, especially perfect and ideal threshold schemes, access structure, and authorized subset. Furthermore, we describe the setup and secret recovering procedures for hash function based schemes, and properties. In Section \ref{ThreatDet} we give reasons for the cyberattack detection model, and then describe it with examples. As integral parts of the model, action database and operation algorithm are introduced with examples. In Section \ref{Setup}, we explain model implementation setup and request, which specify the interconnections of the devices and allowed request types. In Section \ref{implmtn_mehododologies}, we explain the model implementation methodologies that split the task into the frontend UI, backend individual services, and a single database. In Section \ref{demo_usecase}, we demonstrate the implementation on Google Cloud Platform with a use case. Finally, we summarize and conclude the paper in Section \ref{Conclusions}.

\section{Secret sharing schemes}\label{SSS}

A {\bf secret sharing scheme} is a method to distribute a secret among a group of participants by giving a share of the secret to each. The secret can be recovered only if a sufficient number of participants combines their shares.

Formally, we have a secret $S$ and a group of $n$ participants. A {\bf dealer} splits $S$ into $n$ shares and sends shares to each participant privately; then the dealer discards $S$ and all shares. Later on, if a sufficient number of participants returns their shares to the dealer, the dealer combines the shares to recover the secret. Let $t \le n$, an $(t,n)$-{\bf threshold scheme} is one with $n$ total participants and in which any $t$ participants can combine their shares and recover the secret, but not fewer than $t$. The number $t$ is called the {\bf threshold}. Secret sharing scheme was formalized mathematically in independent papers in 1979 by Adi Shamir \cite{SHA79} and George Blakley \cite{BLAK79}. Shamir scheme is based on polynomial interpolation, while Blakley's is from a geometric solution based on hyperplanes. For more detailed discussions, please see our survey article \cite{CFZ18}.

First, we discuss some key concepts and properties of secret sharing schemes. Then we briefly introduce a hash-based secret sharing scheme \cite{CZ13,CZ15} and its properties, which is used as an alternative scheme in the model.  

\subsection{Some key concepts}

A group of participants, which can recover the secret when they return their shares to the dealer, is called an \textbf{authorized subset}. On the other hand, any group of participants that cannot recover the secret is called an \textbf{unauthorized subset}. An \textbf{access structure} $\Gamma$ is a set of all authorized subsets.


Given any access structure $\Gamma$, an authorized subset $A \in \Gamma$ is called a minimal authorized subset if any subset $A' \subsetneq A$ then $A' \notin \Gamma$. We use $\Gamma_0$ to denote a minimal access structure, i.e., the set of all minimal authorized subsets of $\Gamma$. In a $(t, n)$ threshold scheme, let $P$ be the set of the participants, we have

\begin{equation} \Gamma = \{A| A \subseteq P \; \mbox{and} \; |A| \geq
t\}, \;\; \Gamma_0 = \{A| A \subseteq P \; \mbox{and} \; |A| =
t\}. \end{equation}


\noindent In secret sharing, we first define the access structure. Then, we realize the access structure by a secret sharing scheme.


A Shamir $(t, n)$ threshold scheme allows no partial information to be given out even if up to $t-1$ participants joined together \cite{GS97}. A secret sharing scheme with this property is called a \textbf{perfect scheme}. Based on the information theory, the length of any share must be at least as long as the secret itself in order to have perfect secrecy. Up to $t-1$ participants recover no information about the secret under perfect scheme, but when one extra participant joins the group, the secret can be recovered. That means any participant has their share at least as long as the secret. If the shares and the secret come from the same domain, we call it an \textbf{ideal scheme}. In this case, the shares and the secret have the same size.

\subsection{Hash function-based scheme}\label{Subsec_hashSSS}

An efficient and flexible hash function-based secret sharing scheme was designed in \cite{CZ13},  its setup for share generation/distribution and secret recovering are as follows. 

\subsubsection{Setup}

\begin{enumerate}[a)]
\item We randomly generate $n$ distinct shares $s_1, \ldots, s_n$ for $n$
participants $P_1 \ldots, P_n$, where the size of each share is the
same as that of the hash. Then we send $s_i$ to $P_i$ via a secure
channel.

\item We determine all the minimal authorized subsets. Suppose we have
$A_1, \ldots, A_w$ minimal authorized subsets. Each participant
holds a share and combination of the shares of any one of these $w$
authorized subsets will form a private message $M_{priv}$. The
combination will be the concatenation of the shares in participant
sequence. For example, if an authorized subset consists of $P_1,
P_3$ and $P_4$, then the $M_{priv} = s_1 || s_3 || s_4$.

\item Calculate the hash for each authorized subset $A_i$ as
follows

\begin{equation}
H(M_{priv_i}) = h_i,\; i = 1, \ldots, w.
\end{equation}

\item Let $h$ be the secret and of the same size of $h_i$. If we want
the secret to be random, we can generate it in the same way as we do
for shares. Or $h$ is a pre-determined fixed secret.

\item Finally we generate a control $c_i$ for each authorized subset as follows (here
$\oplus$ is bitwise XOR):

\begin{equation}
c_i = h_i \oplus h, \; i = 1, \ldots, w.
\end{equation}

A control also can be used to determine whether a subset is
authorized or not.
\end{enumerate}

After the setup process each participant $P_i$ gets a
random share $s_i, i = 1, \ldots, n$. Public information for each
authorized subset $c_i$, where $i = 1, \ldots, w$, is generated.
Control area $c_i$'s help to herd all the intermediate hashes
$h_i$'s to the final hash $h$. Algorithm \ref{Algo_SSSSetup} shows how secret and shares are being generated. 

\subsubsection{Secret recovering}

Suppose authorized subset $A_i$ consists of participants $P_1, \ldots, P_b$. Joining together they can recover the secret as follows, also see Algorithm~\ref{Algo_SSSRecover}.

\begin{enumerate}[a)]
\item Get the public information $c_i$.
\item $H(s_1 || s_2 || \ldots || s_b) = h_i$, and $h_i
\oplus c_i = h$.
\end{enumerate}

\begin{figure}
\centering
\begin{minipage}{.5\textwidth}
  \centering
      \includegraphics[height=3cm, width=6cm]{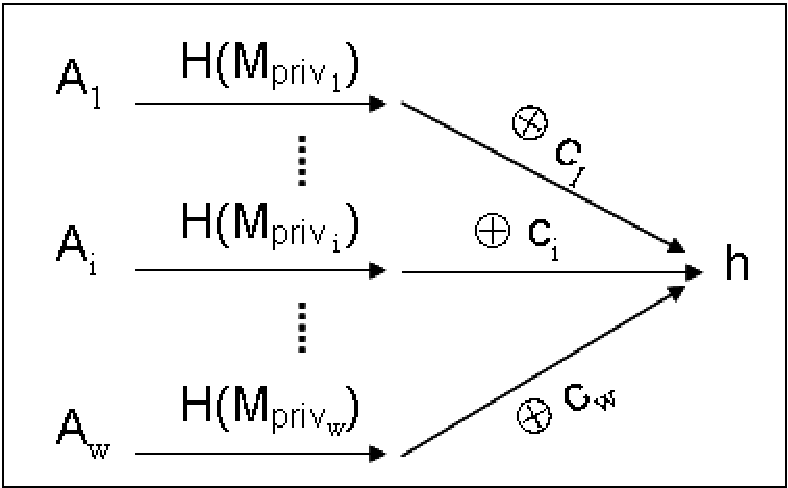}
      \caption{Secret recovery for any authorized subset.}
      \label{fig:SecretRG}
\end{minipage}%
\begin{minipage}{.5\textwidth}
  \centering
     \includegraphics[height=3cm, width=6cm]{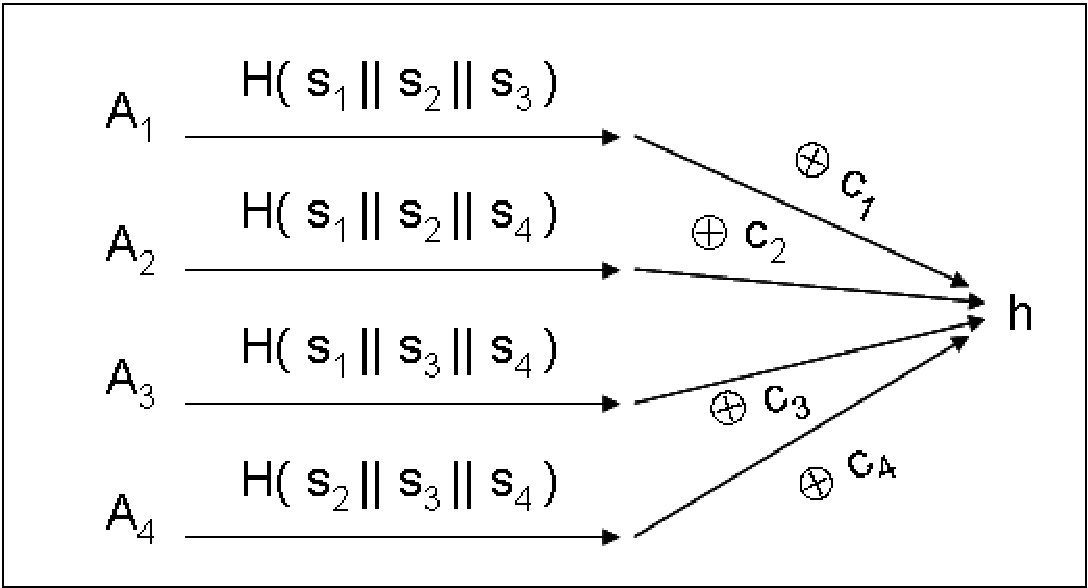}
     \caption{A (3, 4) general threshold scheme.}
     \label{fig:ThreeFourThreshold}
\end{minipage}
\end{figure}

This applies to any authorized subset, see Fig.~\ref{fig:SecretRG} and Fig.~\ref{fig:ThreeFourThreshold}.


\subsubsection{Properties of the scheme}

\begin{enumerate}[a)]

\item We take random bit-strings of the same length for the shares and secret, any group of up to $t - 1$ participants cannot get any additional information about the secret than any outsider. Thus, the hash-based scheme is perfect and ideal. 

\item The calculation of hash function is fast. No complicated or intensive computation, such as polynomial evaluation/interpolation, is needed. The scheme only uses minimal authorized subsets. Thus, we speed up the scheme setup and secret recovery processes. 

\item This approach can be extended to any general access structure. And the scheme is flexible, as any cryptographic hash function can handle any message of arbitrary length so there is no limit to the number of participants.

\item No special hardware or software is required. For example, no need to handle a large number or find a large prime, etc.

\end{enumerate}

\begin{algorithm}
\caption{Function N\_hash\_shares\_setup(threshold, number\_of\_shares)} \label{Algo_SSSSetup}
\KwData{threshold, number\_of\_shares}
\KwResult{secret and a number of shares} 
\DontPrintSemicolon
    \If{threshold $>$ number\_of\_shares}{
        \# fail invalid\;
     }
    \# Create a secret, and the given number of shares, each participant gets one share secret. \;
     array\_of\_shares = create\_crypto\_secrets(number\_of\_shares)

    \# Generate all combinations of participants of size ``threshold'': \;
    \# For example, given tn(2,3), threshold = 2, number of shares = 3, it should return [ [1,2], [1,3], [2,3] ]\;
    list\_of\_authorized\_subsets = generate\_possible\_combinations(number\_of\_shares, threshold)

    \For{each subset in authorized\_subsets}{
       initialize empty private\_message bytes\;
       \For{each participant\_index in subset}{
            \# for example subset = [1,2]\;
            share = get\_shares(participant\_index)\;
            private\_message concatenate share\;

        \# a hashing function of choice, e.g. SHA1, SHA256\;
        hashed\_private\_message = hash\_function(private\_message)\;
        control\_message = hashed\_private\_message XOR secret
      }
    \# associate controlMessage with its Authorized Subset and store safely\;
    store secret, controlMessage and subset indexes\;

    \# each participants gets their own share in later of the process\;
    distribute shares to participants
   }
\end{algorithm}

\begin{algorithm}
\caption{Function recover\_secrets(map\_of\_participant\_index, shares)} \label{Algo_SSSRecover}
\DontPrintSemicolon
    Initialize empty private\_message bytes\;
    \# sort the provided shares by participant index to ensure consistent ordering

   \For{each share in shares orderly} {
       private\_messages concatenate share
    }

    \# a hashing function of choice, e.g. SHA1, SHA256\;
    hashed\_private\_message = hash\_function(private\_message)\;
    control\_message = get control for given participants\;

    \If{control\_message not found or length not matched} {
         \# fail recovery\;
    }
 
    \If{secret == (hashed\_private\_message XOR control\_message)}{
      return successful\;
    }
    \Else{
      return failed\;
    }
\end{algorithm}

\section{The cyberattack detection model}\label{ThreatDet}

\subsection{Model description}

Instead of letting a party own or control a secret, a secret sharing scheme protects the secret by splitting the responsibilities to different shares, each of which is owned by a participant. A secret may be a sensitive document, a decryption key for an important file, or anything such as an index for retrieval of important actions as described below. There are many areas related to secret sharing schemes, but here we want to apply secret sharing to authentication, integration, and detection of unauthorized activities for security improvements.

Nowadays, we are dealing with complex infrastructure and networks consisting of many components (such as servers, routers, switches, computers, smart phones, etc.) linked together. These components communicate with one another and work together to make the whole system functioning. Here we refer to those devices sending and receiving signals as {\bf controllers} and {\bf nodes}, respectively. A signal can be a packet of message, operational control information, or authentication information. Based on software defined network (SDN), communication paths and routes can be easily adjusted to accommodate the traffic conditions at any time. Also, duplicated paths and routes are set up in order to solve problems such as broken links, the single point of failure, etc. We follow the SDN idea to assign more controllers working together for a communication, which is originally done by just one controller. No additional hardware is required. Instead of sending a direct signal by a controller, a few are sending the shares of the signal that are not easily forged \cite{CWZ23}.

\begin{figure}
\centering
\begin{minipage}{.5\textwidth}
  \centering
      \includegraphics[height=3.5cm, width=6cm]{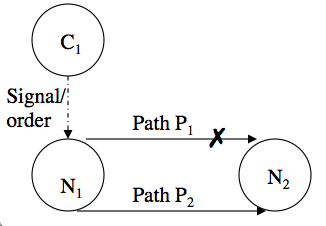}
      \caption{Single controller: switch path to $P_2$.}
      \label{fig:singlecontroller}
\end{minipage}%
\begin{minipage}{.5\textwidth}
  \centering
     \includegraphics[height=4cm, width=6cm]{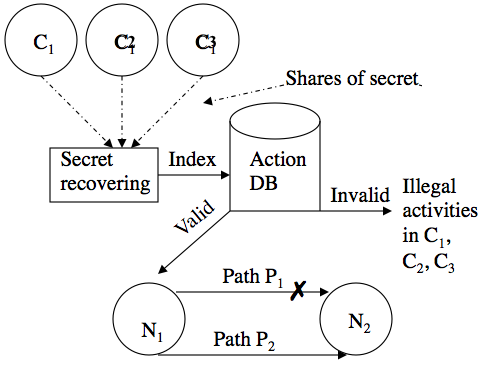}
     \caption{Multiple controllers: switch path to $P_2$.}
     \label{fig:multiplecontroller}
\end{minipage}
\end{figure}

In Fig. \ref{fig:singlecontroller}, suppose controller $C_1$ finds out there is traffic jam in path $P_1$ from node $N_1$ to node $N_2$. It directs/orders node $N_1$ to switch to path $P_2$ to avoid the delays.

What would happen if controller $C_1$ were hacked and under control and the hacker carried out harmful activities to the system? Our solution is to employ more controllers to make the decision for enhanced security. In Fig. \ref{fig:multiplecontroller}, we deploy a $(3, 2)$-threshold secret sharing scheme for authentication. Instead of just sending direct signals, the controllers are sending the shares of secret that are not easily forged if they are well protected. Based on the shares node $N_1$ receives, it tries to recover the secret. In this case, the secret that recovered is just an index pointing to a record in a database which consists of all the actions. The recovered secret or index will be used to search the associated action in the database.  In Fig. \ref{fig:multiplecontroller}, ``switch the path to $P_2$'' will be returned to node $N_1$. If the index is not found in the database, no action will follow, and a detection of illegal activities will come up.

We can see the advantage here. Even if all controllers are compromised, it does not mean that the intruder can get the shares of secrets. This is similar to the analogy that even if a thief can break into a house it does not necessarily mean that he can find the owner's valuables. Also, it is more effective for integration protection and detection. 

\subsection{Action database} 

The action database can be considered as a hierarchy layer for controlling protection measures. Let us use Table~\ref{Example_action_db} to illustrate this. Suppose Node 2 receives a signal from Controller 1. If there is a record with the field Level equal to 0, that means no checking is required for any request from Controller 1 to Node 2. Any request $R_i$ will be allowed if there is a record with field Level 9 corresponding to $R_i$ in the database.

\begin{table}[h]   
    \begin{center}
       \caption{An example action table.}  
       \label{Example_action_db}  
        \begin{tabular}{p{1.2cm}p{1.2cm}p{1.2cm}p{1.2cm}l} 
            \hline
            From & To & Level & Request & Action\\
            \hline
            $\vdots$ & $\vdots$ & $\vdots$ & $\vdots$ & $\vdots$ \\
            1 & 2 & 1 & $R_1$ & $A_1, A_{1a}, A_{1b}$ \\
            1 & 2 & 1 & $R_2$ & $A_2$ \\
            1 & 2 & 1 & $R_3$ & $A_3$ \\
            1 & 2 & 1 & $R_4$ & $A_4$ \\
            1 & 2 & 1 & $R_5$ & $A_5$ \\
            1 & 2 & 2 & $S_6$ & $A_6$ \\
            1 & 2 & 2 & $S_7$ & $A_7$ \\
            1 & 2 & 2 & $S_8$ & $A_8$ \\
            $\vdots$ & $\vdots$ & $\vdots$ & $\vdots$ & $\vdots$ \\
        \end{tabular}
    \end{center}
\end{table}

If there is no record with Level 0 existing, then we continue further checking. Records with field Level equal to 1 are for the first level checking. In Table~\ref{Example_action_db}, there are five Level 1 records. If the request is one of these requests $R_1, \ldots, R_5$, it is valid and allowed to continue process. If it is not, we check if there are any Level 2 records. If no, that means this request is invalid and there will be a feedback to review if Node 1 is compromised. If yes, based on the information about the secret sharing scheme pre-defined in Node 2, we carry out the secret recovering process to recover the secret. If recovered secret is one of the secrets $S_6, \ldots, S_8$, it is valid and allowed to continue. Otherwise, the request will be declined and feedback will be reviewed to see if there is any component(s) being hacked.

For both Level 1 and 2 records, we have an option to append additional requests, just like the original request triggers these additional steps/requests. For example, $A_1$ can trigger $A_{1a}, A_{1b}$, etc. (in Table~\ref{Example_action_db}). This gives us more flexibility. By adding a Level-0 record, we can temporarily halt all the checking in case we have the network upgrade, maintenance, etc.

For those critical requests such as $R_6$, we put $S_6/A_6$ instead of $R_6/A_6$ in the action database. Once node 2 receives $R_6$ request, based on an algorithm and internal information, node 2 will derive $S_6$ and use it to get the corresponding $A_6$. Please refer to \ref{Setup_subsection} for details. 

\subsection{Algorithm}

There are $n$ controllers $C_1, \ldots, C_n$ for a Level 2 security request $R_i$ with corresponding $n$ shares $S_{i, 1}, \ldots, S_{i, n}$. The secret is $S_i$ and the corresponding action in the Action DB is $A_i$. A sufficient number of controllers forms an authorized subset; all authorized subsets form an access structure. The cyberattack detection model runs Algorithm 1 against Action Database to carry out the actions.  

\begin{algorithm}
\caption{Multiple controllers with database} \label{Algo_MultiCtrl}
 \label{Implemt_algorithm}  
 \KwData{Shares from controllers: $S_{i, 1}, \ldots, S_{i, n}$}
 \KwResult{Response from Action Database and/or entries in audit file } 
 \vspace{2mm}

\hspace*{.1em} 10 \hspace{.1em}  Get shares $S_{i, 1}, \ldots, S_{i, n}$ from controllers $C_1, \ldots, C_n$, respectively.\newline
20 \hspace{.1em}  Construct the set of minimal authorized subset $\Gamma_0$.\newline
30 \hspace{.1em}  Pick up an authorized subset $B \in \Gamma_0$ not yet processed. If no more such $B$, respond invalid key, goto 70. \newline 
40 \hspace{.1em}  Recover the secret $Si$ based on $B$. \newline
50 \hspace{.1em}  Use the secret $S_i$ recovered from 40 as key to read the Action DB. If a record $A_i$ is found, goto 60. Else write down $B$ to an audit list file for further analysis, then goto 30. \newline
60 \hspace{.1em}  Carry out the action $A_i$. \newline
70 \hspace{.1em}  Check the audit list file to find out if there are any authorized subset(s) fail to recover the secret. \newline
80 \hspace{.1em}  End.\newline
 
\end{algorithm}



To summarize, we can extend the idea of secret sharing to authentication and cyberattack detection. Instead of relying on one party's decision, we can split it up among a few parties and come up with the decision only if the predefined criteria are satisfied. Also, instead of getting a direct signal, which gets forged more easily, we use the shares of the secret in order to increase the security level.  


\section{Model implementation setup and request}\label{Setup}

We need to specify the interconnections of the devices and what types of requests are allowed. If a request is allowed, depending on the importance, what kind of security measures are needed. This setup is dynamic and needs to be updated whenever required.

Once the setup is done, the system will be functioning accordingly. We refer the messages exchanged between the devices as tickets. It generally contains the required information such as the request or the actions plus the timestamp. One device may receive the same duplicate tickets which are used as consistency and authentication checking.

\subsection{Setup}\label{Setup_subsection}
We have a program $P$, which acts as a security agent to grant access to and communicate with other components and as a dealer to create shares and recover the secret. In Section \ref{implmtn_mehododologies}, program $P$ is embedded in Dealer service. To begin with, we need to create type 9 records in the action database, which are the valid requests. Refer to Table~\ref{Valid_requests}, we have valid requests $R_1, \ldots, R_{99}$ and the corresponding actions $A_1, \ldots, A_{99}$. Then, we need to set up all the communications between controllers/nodes as follow. 

First, $P$ asks if there are any restrictions between the controller and node and then set up accordingly. 

\begin{table}[h]   
    \begin{center}
       \caption{Action table -- valid requests}  
       \label{Valid_requests}  
        \begin{tabular}{p{1.2cm}p{1.2cm}p{1.2cm}p{1.2cm}p{1.2cm}}  
            \hline
            From & To & Level & Request & Action\\
            \hline
            0 & 0 & 9 & $R_1$ & $A_1$ \\
            0 & 0 & 9 & $R_2$ & $A_2$ \\
            0 & 0 & 9 & $R_3$ & $A_3$ \\
            $\vdots$ & $\vdots$ & $\vdots$ & $\vdots$ & $\vdots$ \\
            0 & 0 & 9 & $R_{99}$ & $A_{99}$ \\
        \end{tabular}
    \end{center}
\end{table}

\noindent\underline{\textbf{Case 1: No restrictions}} 

Suppose there is no restriction between controller $C_1$ and node $N_2$. $P$ will create a type 0 record in the action database, see Table~\ref{No_restrictions}. That means $C_1$ can send any request to $N_2$ and as long as there is a corresponding type 9 record in the database, $N_2$ will follow the action in the type 9 record.

\begin{table}[h]    
    \begin{center}
      \caption{No restrictions from $C_1$ to $N_2$}  
      \label{No_restrictions}    
        \begin{tabular}{p{1.2cm}p{1.2cm}p{1.2cm}p{1.2cm}p{1.2cm}}  
            \hline
            From & To & Level & Request & Action\\
            \hline
            1 & 2 & 0 &  &  \\
        \end{tabular}
    \end{center}
\end{table}

\noindent\underline{\textbf{Case 2: With restrictions}} 
\begin{enumerate}[a)]
\item \textbf{Level 1 records:} Suppose there are restrictions between controller $C_3$ and node $N_4$.
$P$ will not create the type 0 record in the action database for $C_3/N_4$. Instead, $P$ will create all the type 1 records corresponding to those requests that are allowed from $C_3$ to $N_4$ in the action database, please see Table~\ref{Restrictions_level_1}. $C_3$ can send requests $R_2, R_7$, and $R_8$ to $N_4$.

\begin{table}[h]  
    \begin{center}
      \caption{Restrictions from $C_3$ to $N_4$ -- level 1 setup }  
      \label{Restrictions_level_1}    
        \begin{tabular}{p{1.2cm}p{1.2cm}p{1.2cm}p{1.2cm}p{1.2cm}}  
            \hline
            From & To & Level & Request & Action\\
            \hline
            3 & 4 & 1 & $R_2$ & $A_2$ \\
            3 & 4 & 1 & $R_7$ & $A_7$ \\
            3 & 4 & 1 & $R_8$ & $A_8$ \\
        \end{tabular}
    \end{center}
\end{table}

\item \textbf{Level 2 records:} $P$ will continue to set up level 2 checking for $C_3/N_4$.
Suppose $R_3$ is in this level, $P$ will create the following type 2 record in the database. Note that $S_3$ is the secret corresponding to $R_3$ in Table~\ref{Restrictions_level_2}. 

\begin{table}[h]  
    \begin{center}
      \caption{Restrictions from $C_3$ to $N_4$ -- level 2 setup }  
      \label{Restrictions_level_2}    
        \begin{tabular}{p{1.2cm}p{1.2cm}p{1.2cm}p{1.2cm}p{1.2cm}}  
            \hline
            From & To & Level & Request & Action\\
            \hline
            3 & 4 & 2 & $S_3$ & $A_3$ \\
        \end{tabular}
    \end{center}
\end{table}

$P$ will determine the participants manually or automatically. Suppose we need a $(2, 3)$ threshold scheme and the participants are controllers $C_3, C_5$ and $C_6$. $P$ runs an algorithm (as dealer) to set up the secret $S_3$ and its shares $S_{3,1}, S_{3,2}, S_{3,3}$ and assign them to $C_3, C_5, C_6$, respectively. $C_3/C_5/C_6$ will save the following information in their internal table, see Table~\ref{Internal_tables}. The first row is the internal table for $C_3$ –- it is the main controller for $C_3$-$N_4$-$R_3$ request and it needs to send out tickets (triggers) to $C_5$ and $C_6$. The second and third rows are the internal tables for $C_5$ and $C_6$, respectively. And $P$ will also create the corresponding share records in the Shares table, see Table~\ref{Shares_tables}. 

\begin{table}[h]  
    \begin{center}
      \caption{Internal tables for three controllers $C_3, C_5, C_6$ }  
      \label{Internal_tables}    
        \begin{tabular}{p{1.2cm}p{1.2cm}p{1.2cm}p{1.8cm}p{1.2cm}l}  
            \hline
            From & To & Request & Participants & Share & Internal Table for\\
            \hline
            3 & 4 & $R_3$ & $C_5, C_6$ & $S_{3, 1}$ & $C_3$ \\
            3 & 4 & $R_3$ &                   & $S_{3, 2}$ & $C_5$ \\
            3 & 4 & $R_3$ &                   & $S_{3, 3}$ & $C_6$ \\            
        \end{tabular}
    \end{center}
\end{table}

\begin{table}[h]  
    \begin{center}
      \caption{Shares table for $P$ }  
      \label{Shares_tables}    
        \begin{tabular}{p{1.2cm}p{1.2cm}p{1.2cm}p{1.8cm}p{1.2cm}}  
            \hline
            From & To & Request & Participants & Scheme \\
            \hline
            3 & 4 & $R_3$ & $C_3$ & $(2, 3)$ \\
            3 & 4 & $R_3$ & $C_5$ & $(2, 3)$ \\
            3 & 4 & $R_3$ & $C_6$ & $(2, 3)$ \\            
        \end{tabular}
    \end{center}
\end{table}

\end{enumerate}

\subsection{Request}

\noindent\underline{\textbf{Case 1:}} When controller $C_1$ sends a request $R_1$ to node $N_2$, $N_2$ will call $P$ to verify.

Since a type 0 record exists for $C_1/N_2$, $P$ will check if there is a type 9 record for $R_1$. If yes, $P$ will send $A_1$ to $N_2$ and acknowledges the approval to $C_1$ and $N_2$. Otherwise, it will inform the decline of the request. \\

\noindent\underline{\textbf{Case 2:}} When controller $C_3$ sends a request $R_2$ to node $N_4$, $N_4$ will call $P$ to verify.

Since a type 0 record does not exist, it will proceed to the next step. Since a type 1 record exists for $C_3/N_4/R_2$, $P$ will send $A_2$ to $N_4$ and acknowledges the approval to $C_3$ and $N_4$. \\

\begin{figure}
\centering
\includegraphics[height=21cm, width=14cm]{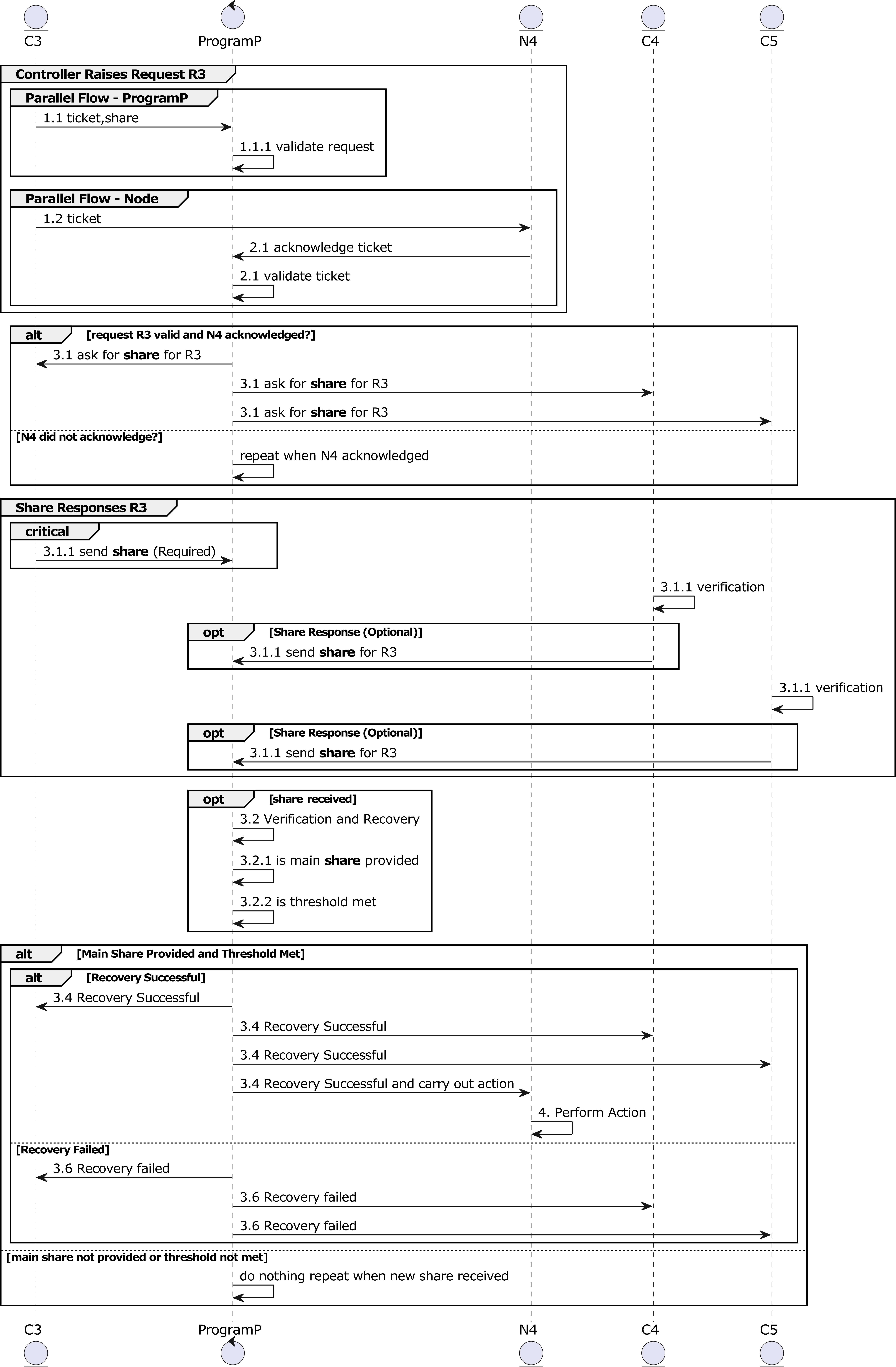}
\caption{Request case 3 model sequence in a diagram.} \label{fig:Request_case3}
\end{figure}

\noindent\underline{\textbf{Case 3:}} Controller $C_3$ sends a request $R_3$ to node $N_4$, see Fig.~\ref{fig:Request_case3} for request sequence diagram.

When $C_3$ sends a request $R_3$ to $N_4$, it finds out $C_3$-$N_4$-$R_3$ is in its internal table. $C_3$ sends to $P$ and $N_4$, then $N_4$ needs to acknowledge the ticket. When $P$ has both the initial request from $C_3$ and the acknowledgement from $N_4$, it verifies request shares from participants. It also sends out the ticket and its share to $P$. $N_4$ will send out the ticket to $P$ for approval. Once $P$ verifies the consistency of the tickets from $C_3$ and $N_4$ and finds out $C_3$-$N_4$-$R_3$ is in the shares table, it generates a ticket and sends it to $C_5, C_6$ to ask for their shares.

Controllers $C_5, C_6$ will verify the consistency of the tickets from $C_3$ and $P$. If yes, they will send their tickets and shares $S_{3,2}$ and $S_{3,3}$ to $P$, respectively.  

After verification, $P$ recovers the secret $S_3$ using $(S_{3,1}, S_{3,2})$ or $(S_{3,1}, S_{3,3})$ and gets $A_3$ from the action database. [Note that we use $(S_{3,1}, S_{3,2})$ or $(S_{3,1}, S_{3,3})$ as minimal authorized subsets because $C_3$ is the main controller, we emphasize its importance/contribution when recovering the secret, and ignore the authorized subset $(S_{3,2}, S_{3,3})$.] It sends the actions(s) $A_3$ to node $N_4$ together with the ticket. $P$ will inform $C_3$ and $N_4$ if the request is approved. $P$ also sends ``approved'' to $C_5$ and $C_6$, so it can mark for UI to grey out. \\

\noindent {\bf{Steps for $C_3$:}} 
\begin{enumerate}[$\cdot$]
\item {\bf{1.1}}: Timestamp-$C_3$-$N_4$-$R_3$, $S_{3,1}$  (ticket and share to $P$)
\item {\bf{1.2}}: Ticket to $N_4$
\end{enumerate}

\noindent {\bf{Steps for $N_4$:}} 
\begin{enumerate}[$\cdot$]
\item {\bf{2.1}}: Timestamp-$C_3$-$N_4$-$R_3$  (ticket to $P$)
\end{enumerate}

\noindent {\bf{Steps for $P, C_5, C_6$:}} 
\begin{enumerate}[$\cdot$]
\item {\bf{3.1}}: Ask for shares form $C_5$ and $C_6$

\begin{enumerate}[$\cdot$] 
\item {\bf{3.1.1}}: Verification the consistency for the tickets both from $C_3$ and $P$
\item {\bf{3.1.2}}: Timestamp-$C_3$-$N_4$-$R_3$, $S_{3,2}$ (ticket and share to $P$ from $C_5$)
\item {\bf{3.1.3}}: Timestamp-$C_3$-$N_4$-$R_3$, $S_{3,3}$ (ticket and share to $P$ from $C_6$)
\end{enumerate}

\item {\bf{3.2}}: Verification for consistency of tickets 
\item If yes from {\bf{3.2}}, do {\bf{3.3}}, {\bf{3.4}} and {\bf{3.5}}
\item $\;\;\;$ {\bf{3.3}}: Recover secret $S_3$ and get action(s) $A_3$ 
\item $\;\;\;$ {\bf{3.4}}: Send action(s) $A_3$ to $N_4$
\item $\;\;\;$ {\bf{3.5}}: Inform approval of request $C_3$-$N_4$-$R_3$ to $C_3$ and $N_4$ 
\item If no from {\bf{3.2}}, do {\bf{3.6}}
\item $\;\;\;$ {\bf{3.6}}: Send reject signal to $C_3$ and $C_4$
\end{enumerate}

\noindent {\bf{Steps for $N_4$:}} 
\begin{enumerate}[$\cdot$]
\item {\bf{4.1}}: Once there is a match between the tickets from $C_3$ and $P$, carry out actions(s) $A_3$ 
\end{enumerate}


\section{Model implementation methodologies}\label{implmtn_mehododologies}


The model implementation is split into three sections, the UI also known as the frontend, individual services also known as the backend, and a persistence layer using a single Postgres Database.

\subsection[short]{UI - Frontend}


The UI operates on an HTTP server fulfilling two primary purposes. Its first purpose is to deliver the UI dashboard page to end-users accessing it via a browser. This dashboard, constructed with HTML and Javascript utilizing React.js and Bootstrap for component development, serves as the main interaction point for users to run project simulations. It displays and allows on-demand editing of scheme configurations, shares, action requests, and route configurations to control overall system behavior.


The UI's second purpose is to provide endpoints that direct calls from the dashboard to the appropriate backend services. This architecture eliminates the need for the UI dashboard to know the individual URLs of each service, instead requiring it only to call its service. For instance, when a user updates the secret share scheme to a hash function-based scheme and adjusts its threshold, the dashboard sends an HTTP POST request to the UI server. The UI server then uses a Naming Registry to locate the dealer's URL and forwards the request to the dealer accordingly.

\subsection[short]{Python - Backend}


The model's backend implementation is written in Python and comprises four main services.

\begin{enumerate}[1)]
\item Dealer - Responsible for creating, distributing shares, and recovering secret
\item Controller - Holding on one piece of share, requesting actions and responding to other controllers requests
\item Node - Performing actions upon a successful secret recovery
\item Name Registry - A registry of service URLs and directing traffics
\end{enumerate}

\subsubsection{Core Framework}


For all of the services, we can think of it as a micro-service architecture, where each individual service is designed with specific, distinct responsibilities. To maintain consistency across these independent services, this core framework is developed to handle commonly used functions. For communication between services, each service has implemented a set of RESTful endpoints. Each endpoint allows the service either accepting an input via HTTP POST request, or providing information via HTTP GET request. This project leverages the Twisted [https://docs.twistedmatrix.com] as the core library to run each service as an HTTP server individually. Below is a list of the Python libraries used in the implementation.

\begin{itemize}
\item \textbf{requests} - makes HTTP requests
\item \textbf{psycopg-binary} - PostgreSQL Driver for Python
\item \textbf{psycopg} - PostgreSQL Driver Python wrapper
\item \textbf{twisted} - HTTP server library
\item \textbf{shamirs} - A Python implementation of Shamir scheme
\end{itemize}

\subsubsection{Endpoints for Simulation}


To facilitate end-user simulation of various use cases, all services include a set of administrative endpoints designed to update their internal data. For instance, a dedicated `reset' endpoint on a controller service allows resetting all shares, while a similar `reset' endpoint on a dealer service resets all action and action request information. Additionally, endpoints prefixed with 'command' enable end-users to simulate and trigger specific actions on a service.

\subsubsection{Dealer}
A dealer service is responsible for generating and distributing shares to participating controllers, performing secret recovery for an action request when the required number of shares are collected, and auditing logs.

First the dealer service has a set of endpoints that are related to scheme and action configurations. One endpoint facilitates the configuration of a \textbf{Secret sharing scheme}, accepting either a \textbf{Hash-based scheme} or a \textbf{Shamir scheme}. This configuration includes parameters for threshold and total shares to be generated based on the number of participants. Another endpoint is implemented to accept a list of actions from an admin user. The list of actions should have a \textbf{from} controller, a \textbf{to} target node, the security \textbf{level} the \textbf{request} name or identifier, and \textbf{action} alias, see Table~\ref{Example_action_db} for an example. A third endpoint is to trigger the secret generation. It starts processing one row at a time. For each row it will generate secret and create shares according to the number of participants, then distribute them.


The next set of endpoints are related to secret recovery. One endpoint accepts an action request from controllers. Another endpoint accepts a share during an action request that a participant controller responds the request its share. This endpoint collects the shares, when the number of shares meet the threshold, it triggers the secret recovery function. There are two outcomes from the secret recovery function, passed or failed. In both case, the dealer notifies all involved controllers, but only notifies node when it is passed.

\begin{itemize}
\item \textit{/command/action} allows an end-user to send a list of request and action configurations to generate secrets and shares.
\item \textit{/command/init-action} allows an end-user to trigger the secret and share generation and distribution.
\item \textit{/command/scheme-config} allows an end-user to configure the scheme and threshold values.
\item \textit{/command/tn-participant-config} allows an end-user to choose the participants and selection strategy when deciding which controller to pick when generating shares.
\item \textit{/accept-shares} accept share from a controller.
\item \textit{/action-request} accepts a new action request from a controller.
\item \textit{/forward-action-request} accepts an acknowledgment from a node for an action request, then asks the participant controllers to return the shares.
\end{itemize}

\subsubsection{Controller}

A controller service is responsible for holding on the shares of any given action that it is participated in, raising action requests, and responding the share when an action request needs it. For this model implementation we also simulate that a controller is corrupted or comprised by altering any of the share so that secret recovery is not successful. The following is the list of endpoints and their behaviors. 

\begin{itemize}
  \item \textit{/secret} allows dealer to push shares to the controller. This also allows the end-user to simulate a corrupted controller by sending in a new share.
  \item \textit{/command/action-request} allows the end-user to simulate raising an action-request of any action if this controller is the main controller.
  \item \textit{/ask-for-shared-secret} raises an action request, the dealer is asking participated controllers to respond its share for this action.
  \item \textit{/command/respond-share} allows, for simulation, an end-user to trigger the controller to react to an action request on demand.
  \item \textit{/action-request-result} allows the dealer to push secret recovery result to the controller.

\end{itemize}

\subsubsection{Node}

A node is a service that carries out an actual action. Currently, upon an action request sent from a controller,  the corresponding  node is implemented to automatically respond with an acknowledge message to the dealer. This acknowledgment will allow the dealer try to recover the secret when the required number of shares are collected. On the other hand, it would perform an actual action when the dealer sends a successful recovery message for a given action request. Each of these is an HTTP endpoint that accepts valid messages.

\begin{itemize}
\item \textit{/action-request} accepts an action-request then automatically responds to the dealer that it acknowledges it.
\item \textit{/action-request-result} when a secret recovery is successful, the node will perform the action.   
\end{itemize}  

\subsubsection{Name Registry}
One of the crucial components in this project is the Name Registry service. It acts as a DNS (domain name system) provider, when a service name is queried, it would return its URL. However, because of a routing requirement a more advanced implementation is built into the Name Registry. Out of all messaging between individual services, either it is a share request, or a recovery response, or an action, the main goal is to guarantee the message delivery. When a path is broken between service A and service B, the routing mechanism would look for an alternate route, maybe from service A to Service C, then from Service C to service B to deliver this message. On a second scenario, if the path from service A to service B has high traffics or weights, then the routing mechanism would provide a better route to deliver the message.

Therefore, Name Registry has all URLs on all services and the weights between them. These weights are editable to simulate when a link is broken, or traffic increases between any of the nodes. To achieve the requirement in a more flexible way, it is implemented as a turn by turn basis, similar to a GPS alike behavior. In the following explanation, we would use the term, origin, destination, from-node, and to-node to describe each party for a given message to deliver. The Name Registry itself contains some computation for any given hosts using Dijkstra's shortest path. For example, when a message is originated from service A, and the destination is Service B. The condition of the network or paths is that, Service A has no direct connection to Service B. In this case, service A would first asks Name Registry, that it would want to go to service B (destination). The name registry computes a shortest path, and in this example, we would say Service C is one unit(weight) to Service A, and also one unit to Service B. It is obvious that From A to B, it can first go to C, then B. There for, Name Registry would return the URL of service C. Then Service A would put all the information of origin (Service A), destination (Service B), from-node (Service A), and to-node (Service C) into the message then send it over to Service C. This is because we wanted to achieve a turn by turn lookup, so that, it is more real time and can adapt sudden changes to make additional detours when needed.
Service C now received the message, and it looks at the destination information and it is not for Service C. So it would ask Name Registry for the routes from service C to service B. And for this request, the URL of Service B is returned. Service C would then package a similar destination information like Service A, and send the message. Finally when Service B gets the message, it checks and it is for service B, it continues to handle specific behavior for the message as normal.

\begin{itemize}
  \item \textit{/route} allows an admin user to configure the path between services to simulate broken links or heavy traffic paths.
  \item \textit{/path} allows a service to query for destination and to return the next stop URL. 
\end{itemize}

\subsection[short]{Database}

The choice of database in this model implementation is PostgreSQL. It is free and ACID (atomicity, consistency, isolation, and durability) compliant and used to stored actions and logs for recovery attempts. The following is the table schema.

Table: action \\

Columns:
\begin{itemize}
\item from\_node
\item to\_node
\item level
\item request
\item action
\end{itemize}

Table: audit\_recovery \\
Columns:
\begin{itemize}
\item    id       serial primary key
\item    reference\_number 
\item    request          
\item    batch\_id        
\item    is\_success       boolean,      not null
\item    context\_nodes    extra context information about the recovery
\item    created\_at       timestamp
\end{itemize}

\subsection{Running Environment}

As previously stated, the model consists of four services, \textbf{Dealer}, \textbf{Controller}, \textbf{Node} and \textbf{Name Registry}. Out of the four services, \textbf{Dealer} and \textbf{Name Registry} have only one running instance each. However, \textbf{Controller} and \textbf{Node} in theory can have unlimited number of instances running. For demonstration purpose, we should have at least three \textbf{Controller} instances and two \textbf{Node} instances running in a complete system. This should relates to the threshold and total number of participants configuration of (\textit{t},\textit{n}) threshold scheme. With this minimal system setting, we are to manage seven HTTP servers and one database. Docker [https://docker.com] is used to help managing the start and stop of individual services and their URLs.

A docker compose yaml file contains all necessary information, including the number of \textbf{Controllers} and \textbf{Nodes} and their CPU and memory settings. Initializing the PostgreSQL database with the DDL (data definition language) schema and ensuring the services can connect to the \textbf{Name Registry} during run time. Finally, another key configuration is an ini file that maps the name of the service instances to the URLs. This file is loaded into \textbf{Name Registry} after it starts up. Then \textbf{Name Registry} would distribute the \textbf{names} to all instances. When the instance queries for a destination, the instance needs to send its \textbf{name} as its identifier to help computing the path to destination. The ini file is not the optimal solution because it requires manual effort to denote how many instances, their URLs, and Name Services restarted to take effect. However, it is a suitable solution as opposed to a more complex services discovery solution. By swapping ini configurations, the complete system can be run on the OS level of any personal computer, that is beneficial for local development of the model implementation. It can also run and scale up on controllers and nodes on docker, that is beneficial in running automated tests on a complete system.

Another reason of using docker is that it provides the best portability and consistency on any platform. During build time, all Python scripts and required libraries are packaged using a \textbf{Debian 12 Bookworm} [https://www.debian.org/releases/bookworm/] with \textbf{Python 3.12} image. Therefore, it can run on any underlying platform, operating system and CPU architecture as long as docker is supported.

\textbf{Google Cloud Platform} provides a Virtual Machine with more than enough resources, and the model is running six \textbf{Controllers}, six \textbf{Nodes}, one \textbf{Dealer}, one \textbf{Name Registry}, and finally one database. This setup on Google Cloud Platform is large enough to run complex scenarios but still be manageable for navigating UI dashboard to perform any end-user simulations.

\section{Demo user interface and use cases}\label{demo_usecase}

To illustrate the idea, we implemented the proposed model on Google Cloud Platform with graphic user interface. In this section, we will explain and demonstrate how to use the user interface (UI) to simulate a complete flow of action requests and secret sharing schemes. Assuming there are a running instance of the UI, some controllers and nodes, and one dealer service. The demo is accessible by this link http://www.techep.csi.cuny.edu, on a server located in Computer Science Department at College of Staten Island.

\subsection{UI overview}\label{sec:ui-overview}

    The UI is broken up into sections, and each represents a different part of the system, see Fig.~\ref{fig:UI_overview}. The sections are as follows:
  \begin{figure}  
    \begin{center}
        \includegraphics[scale=0.13]{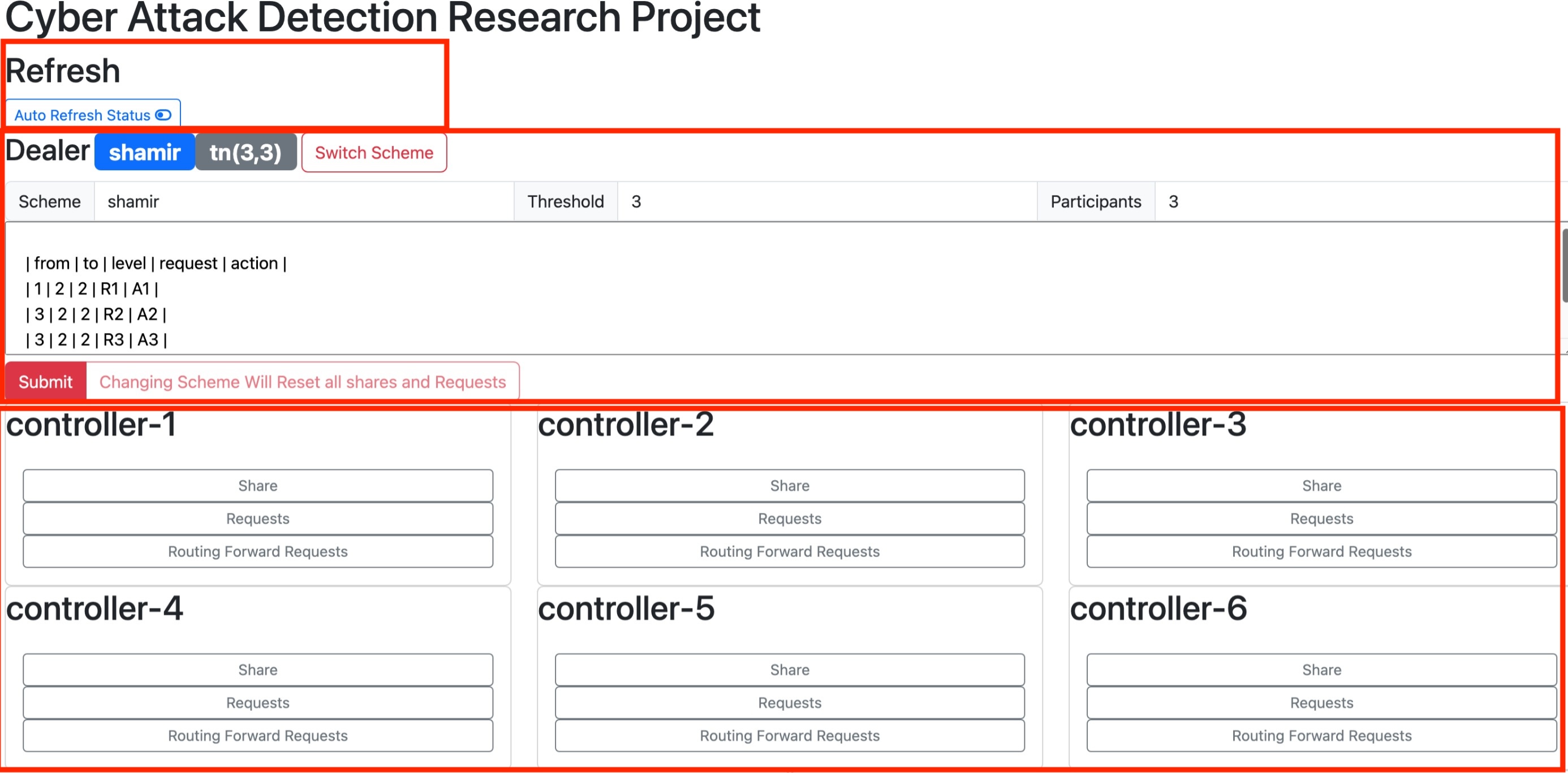}
        \includegraphics[scale=0.13]{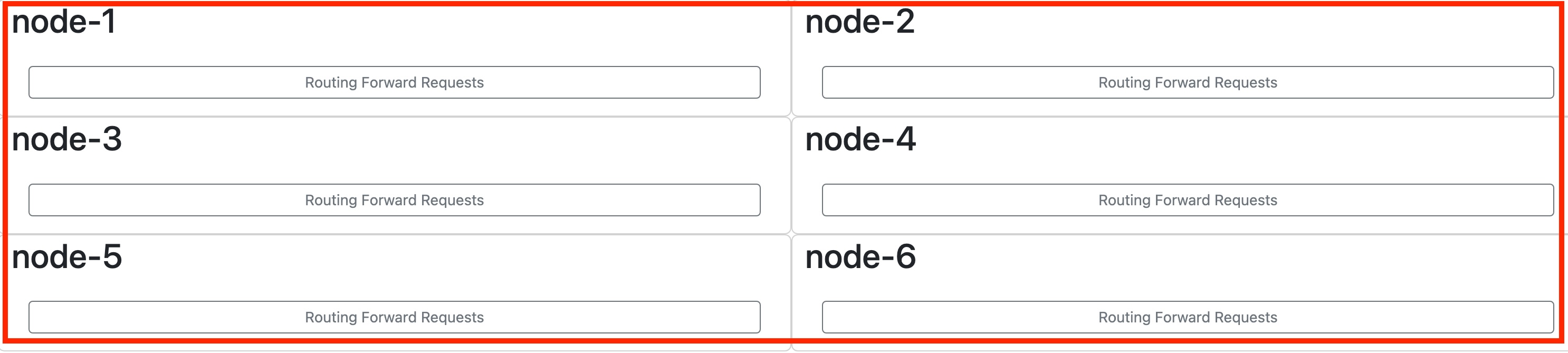}
        \includegraphics[scale=0.13]{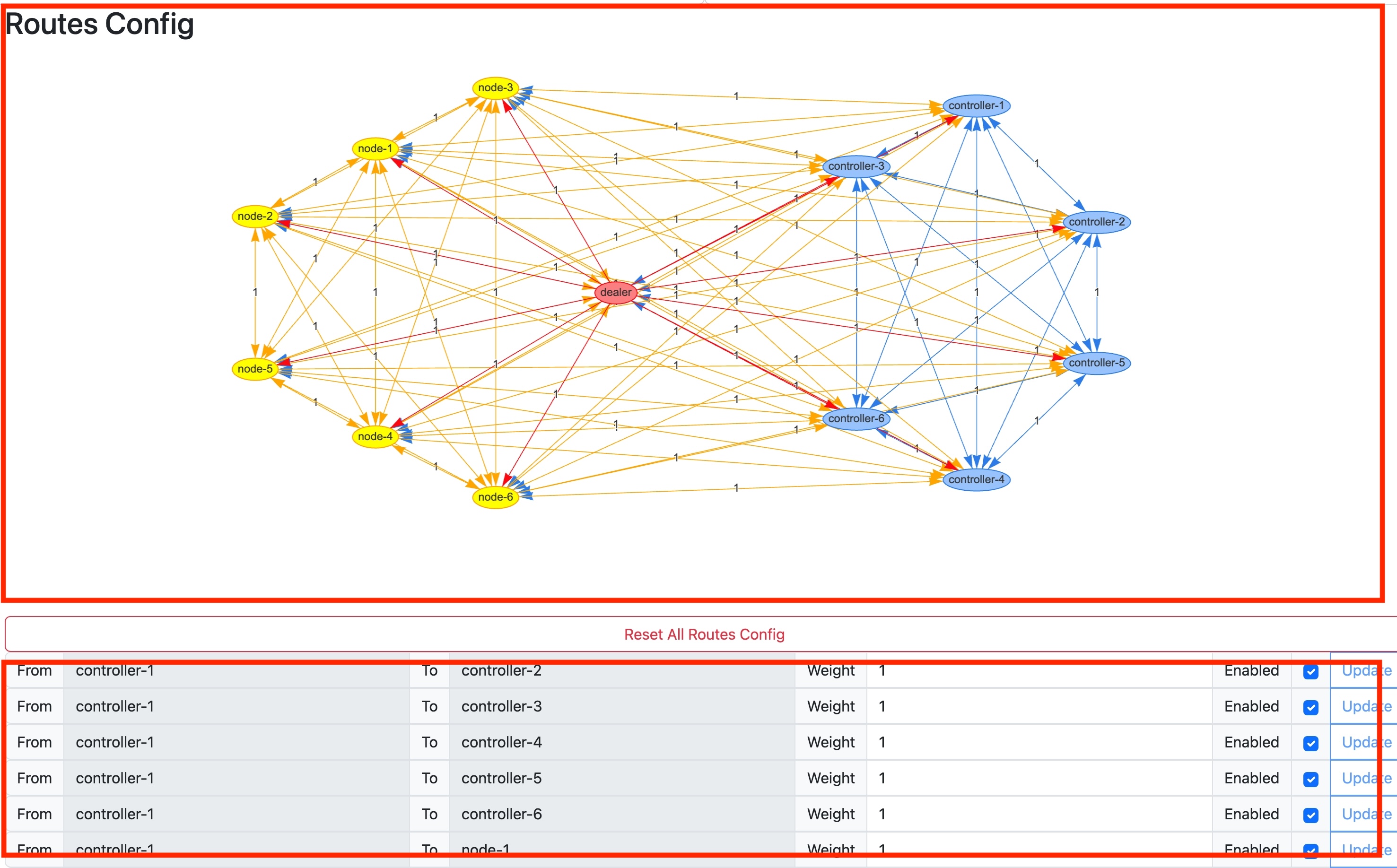}
        \caption{Secret sharing scheme based cyber attack detection project graphic user interface.} \label{fig:UI_overview}
    \end{center}
   \end{figure}
    
    \begin{itemize}

        \item \textbf{Refresh} - The refresh button allows page auto-refresh every five seconds when it is turned on (click/press the button to make sure the little blue dot is on the right side).

        \item \textbf{Dealer Service} - The dealer service is responsible for accepting inputs to generate shares for a secret.

        \item \textbf{Controllers} - A list of all the controllers in the system. 
        Each controller has its own controllers and buttons to raise requests, respond to requests, and view the status of the requests. 
        It is also possible to alter the share of a controller to mimic the situation of a compromised controller.

        \item \textbf{Nodes} - A list of all the nodes in the system.
        Each node has its own area to display requests that are sent to this node, and the event logs of the request.
        And lastly if a request is approved, the node will carry out the action and display the result.

        \item \textbf{Routes Config} - I.e., the routes configuration, allows the user to edit the links or paths between the services. By default, all services (Controller, Nodes, and dealer) are connected to each other with a weight or distance of one.
    \end{itemize}

\subsection{Dealer service}\label{sec:ui-dealerservice}

    \enumerate

\begin{figure}
    \begin{center}  
        \includegraphics[scale=0.35]{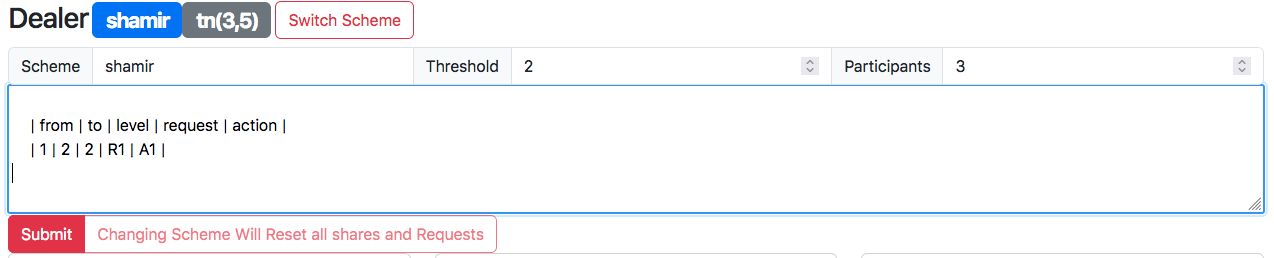}
    \caption{Dealer service configuration and editing action database record}\label{fig:UI_dealerservice}
    \end{center}
\end{figure}

    \item By clicking ``Scheme'' box, we can configure the dealer to use \textbf{Shamir scheme} by choosing ``Shamir'' or \textbf{Hash-based Scheme} by choosing ``Hash.'' They both accept a threshold configuration of \textbf{$(t, n)$}, where \textbf{$t$} is the threshold value, \textbf{$n$} is the total number of controllers (participants) to distribute the shares. For example in Fig.~\ref{fig:UI_dealerservice}, by choosing ``Shamir'' in ``Scheme'' box, entering 3 in ``Threshold'' box and 5 in ``Participants'' box, we configure the system to use \textbf{$(2, 3)$-threshold Shamir secret sharing scheme}.  

    \item Below is an editor area (if it is collapsed/folded, click the \textcolor{red}{ ``Switch Scheme''} button to expand/open the area), that allows user to enter the Action Database record delimited by a vertical bar `\textbf{$|$.}' For example in Fig.~\ref{fig:UI_dealerservice}, the line in the edit area would signal the dealer, it is a Level 2 security request \textbf{$R_1$} from Controller \textbf{$C_1$} to Node \textbf{$N_2$}, and the action is \textbf{$A_1$}. Indeed, what happens behind the scene is that the dealer first recovers the secret $S_1$, then uses $S_1$ as the key/pointer to map to the request $R_1$. Upon receiving the payload, the dealer should generate the secrets and shares based on scheme configuration. In this example, it is using \textbf{$(2, 3)$-threshold Shamir scheme}, and the Controller \textbf{$C_1$} is the \textbf{main} participant amongst the other two. Once finishing the editing of the Action Database records and selection of secret sharing scheme (either Shamir or Hash-based), click the \textcolor{red}{red} \textbf{Submit} button to reset all shares and requests. 

    Below is an example of a configuration with four requests entered in the editor area of the dealer service, $R_1, R_2, R_3, R_4$, and the corresponding actions $A_1, A_2, A_3, A_4$. Controller $C_1$ is the main participant for request $R_1$, Controller $C_3$ is the main participant for $R_2$ and $R_3$, and Controller $C_2$ is the main participant for $R_4$. 

    \begin{lstlisting}[mathescape][label={lst:lstlisting2}]
       | From | To | Level | Request | Action |
       | 1    | 2  | 2     | $R_1$        | $A_1$       |
       | 3    | 2  | 2     | $R_2$        | $A_2$       |
       | 3    | 2  | 2     | $R_3$        | $A_3$       |
       | 2    | 1  | 2     | $R_4$        | $A_4$       |
    \end{lstlisting}

    \endenumerate

\subsection{Controllers}

\begin{figure}
    \begin{center}
        \includegraphics[scale=0.35]{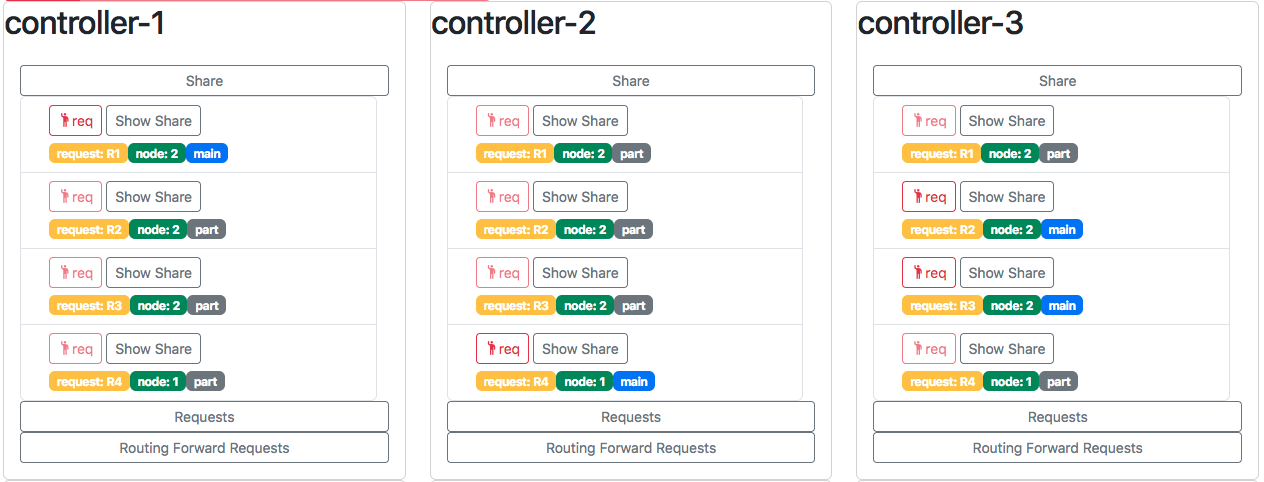}
        \caption{Share areas of controllers $C_1, C_2, C_3$ before issuing any request.}\label{fig:UI_b4request}
    \end{center}
 \end{figure}

    The controller section displays more than one controller, and it wraps to next row if it cannot fit. In this demo, we use up to six controllers, see Fig.~\ref{fig:UI_overview}. Each controller has three expandable areas: Share, Requests, and Routing Forward Requests. Based on the example configuration at the end of section~\ref{sec:ui-dealerservice}, after pressing the ``Submit'' button in the Dealer section of the UI, the controllers $C_1, C_2$, and $C_3$ display four requests in their ``Share'' areas as in Fig.~\ref{fig:UI_b4request}.   
    
\enumerate

    \item \textbf{Share} \\ 
    This section shows all shares that this controller has in memory along with their request information. By clicking and expanding ``Share'' area, we see an icon of a little man with the right hand raised followed by ``req'' button (raise-hand) and ``Show Share'' button. \newline
   
\begin{figure}
    \begin{center}
        \includegraphics[scale=0.5]{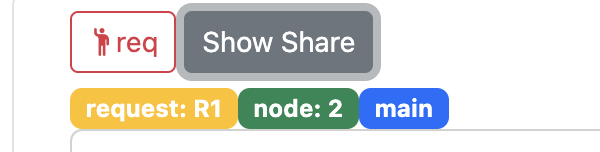}
         \caption{Raise-hand button for controller to raise request.}\label{fig:Raishand}
    \end{center}
 \end{figure}

    Raise-hand \textbf{req} button: If the controller is the main participant for the request, then the raise-hand button in \textcolor{red}{\textbf{bold red}} is enabled and can be clicked, see Fig.~\ref{fig:Raishand}. Otherwise, the button will be grayed in \textcolor{red}{light red} and cannot be clicked, which means the the controller in question is not the main participant and cannot raise request. Only main participant can raise requests, clicking on the raise-hand button will send the request to the dealer service. In Fig.~\ref{fig:UI_b4request}, the controller $C_1$ raise-hand \textbf{req} button for request $R_1$ is highlighted below. It also shows that the destination node for this request is node $N_2$ and $C_1$ is the main participant.    

\begin{figure}
    \begin{center}
        \includegraphics[scale=0.35]{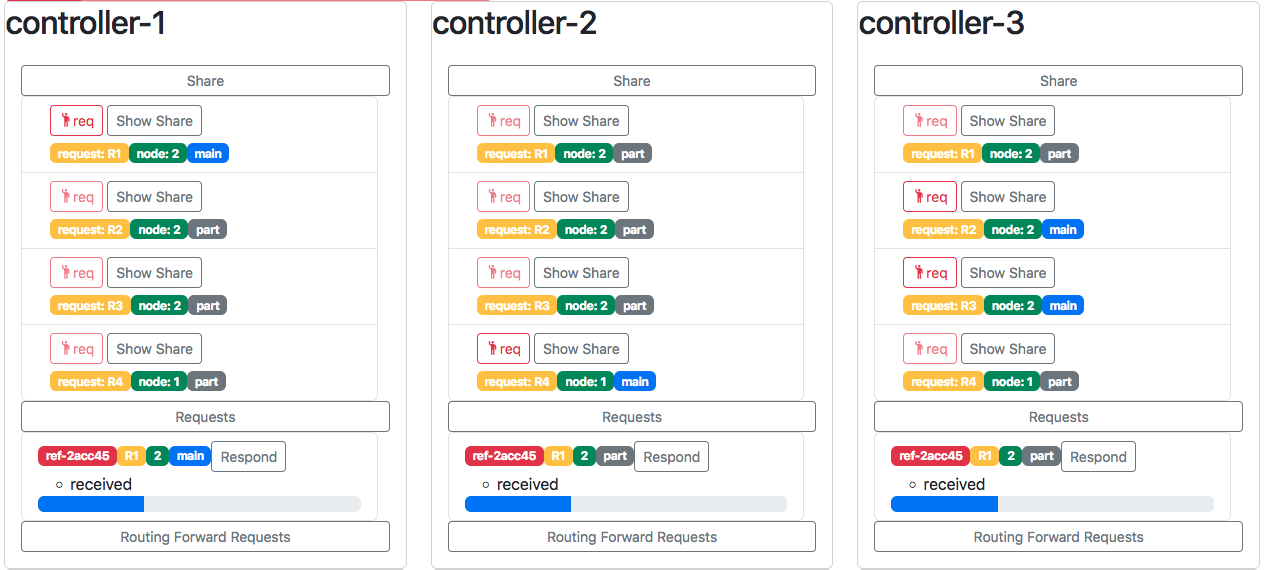}
        \caption{Share areas of controllers $C_1, C_2, C_3$ after raising request $R_1$.}\label{fig:UI_afterrequest}
    \end{center}
 \end{figure}

After raising request $R_1$ by clicking \textbf{req} button, Fig.~\ref{fig:UI_b4request} becomes Fig.~\ref{fig:UI_afterrequest}, where the request $R_1$ is received by three participating controllers $C_1, C_2$, and $C_3$. 
    
\begin{figure}
    \begin{center}
        \includegraphics[scale=0.5]{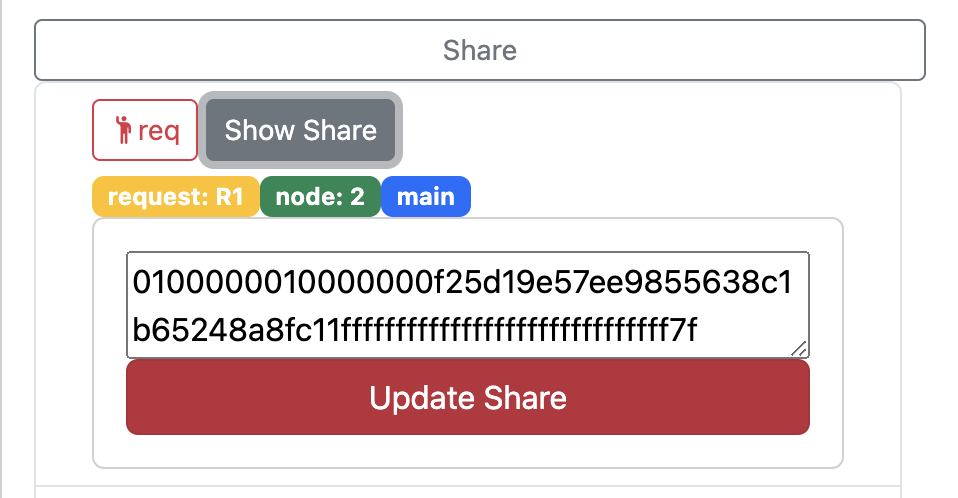}
        \caption{Change share to mimic a compromised controller.}\label{fig:ChangeShare}
    \end{center}
 \end{figure}

   \textbf{Change Share} 
    Clicking open \textbf{Share} button will show the \textbf{share} associated with this request. The user can change this value then click \textbf{Update Share} button at the bottom to change the value, see Fig. \ref{fig:ChangeShare}. Note that this action is unrecoverable, and it mimics the situation of a compromised controller.

    \item \textbf{Request Status} \\ 
    This section shows all requests that are currently in the system. It shows the status of the request, the action that is associated with the request, and the request number. The controllers that are part of a request will have a \textbf{Respond} button to respond to the request.
    Clicking on the button will signal the controller to send its share to dealer for this request for dealer secret recovery. \\
    Status indicator lines will display whether the request is successfully recovered or failed
    \enditemize

    \item \textbf{Routing Forward Request} \\ 
    This section shows all ``passthrough'' requests that are for another controller. Due to higher routing cost, the requests are routed via this controller because of a shorter path.

\endenumerate

\subsection{Nodes}

\begin{figure}
    \begin{center}
        \includegraphics[scale=0.3]{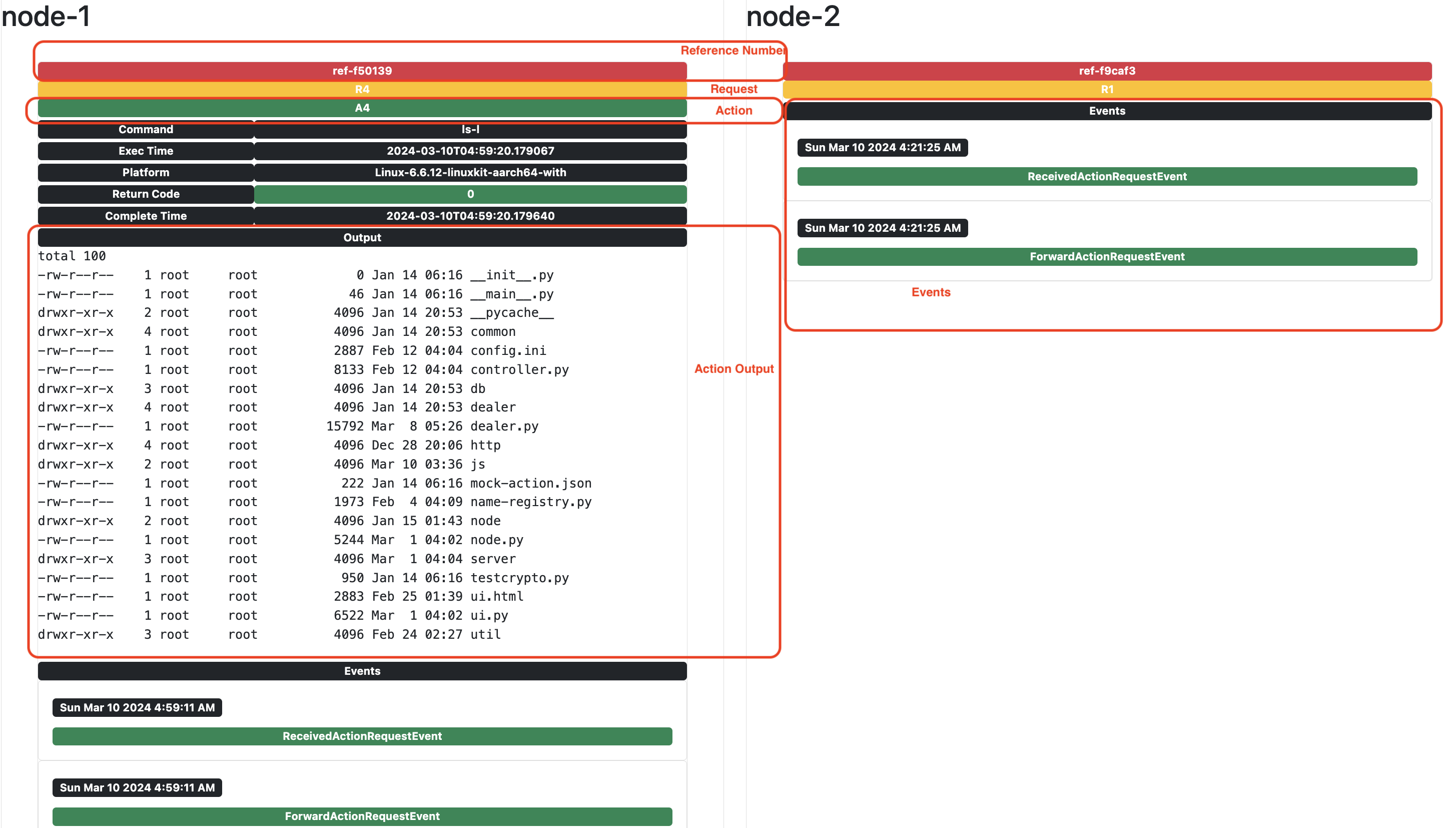}
        \caption{Display of node section.}\label{fig:NodeSection}
    \end{center}
 \end{figure}

    The nodes section displays more than one node, and it wraps to next row if it cannot fit, see Fig. \ref{fig:NodeSection}.

    Each node has an area that shows
    \itemize
    \item \textbf{Request information} \\  
    The reference number, request name, etc.

    \item \textbf{Action information} \\  
    When the request is approved, the node will perform the action.
    Most of these are system commands (e.g. shell commands like \textit{ls}, \textit{pwd}, etc.). The execution status and results are displayed here.

    \item \textbf{Event logs} \\  
    The journey of the requests. The node will start logging events when it receives the request. Upon receiving a recovery success, the node will carry out the action and display the result. Each step will have its corresponding log and timestamp.

    \item \textbf{Routing forward request} \\  
    Similar to controllers, each node also has a routing forward request section. This section shows all ``passthrough'' requests that are for another node. 

\begin{figure}
    \begin{center}
        \includegraphics[scale=0.13]{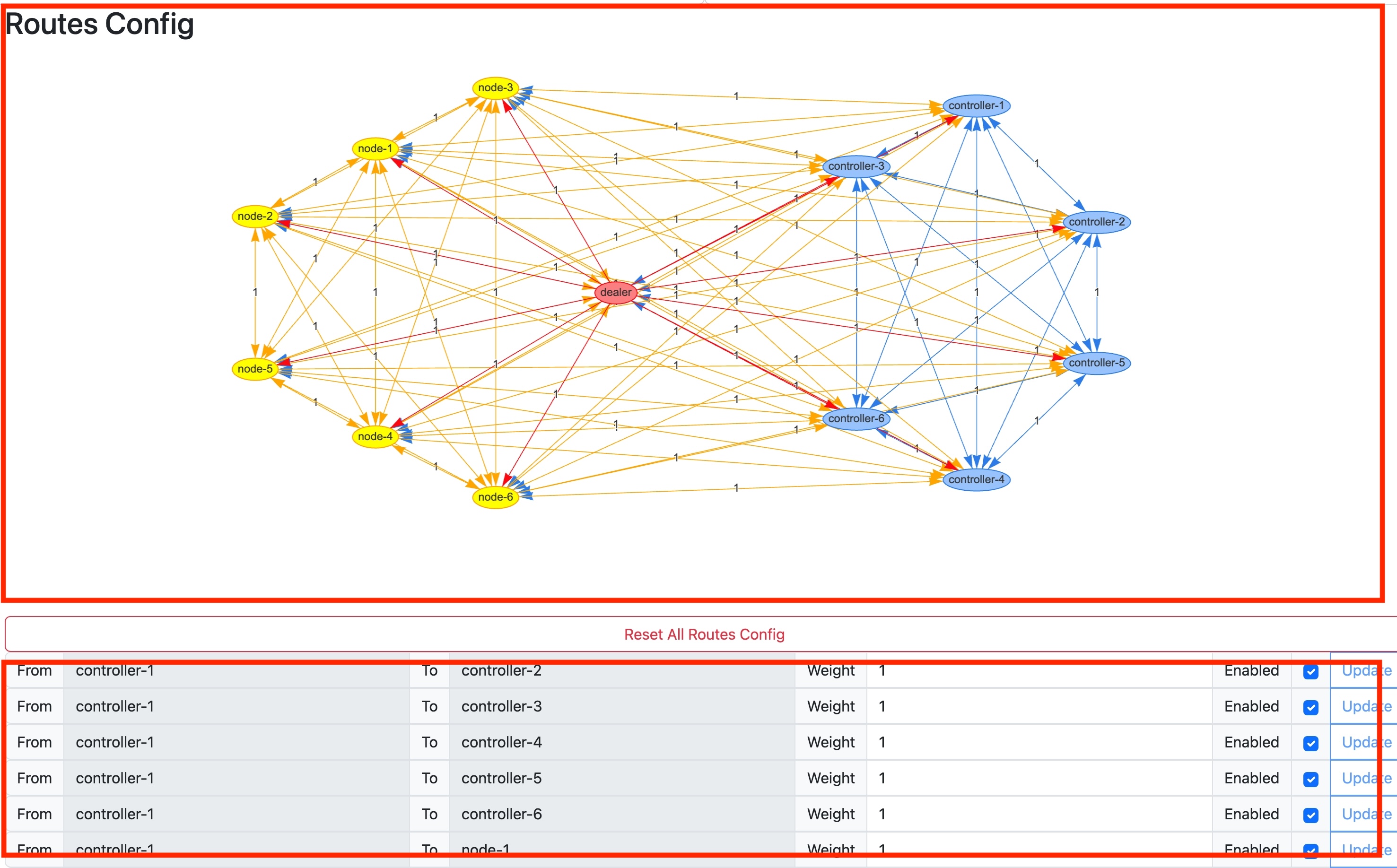}
        \caption{Routes configuration UI and default weights of links.}\label{fig:RoutesConfigUI}
    \end{center}
 \end{figure}

\begin{figure}
    \begin{center}
        \includegraphics[scale=0.23]{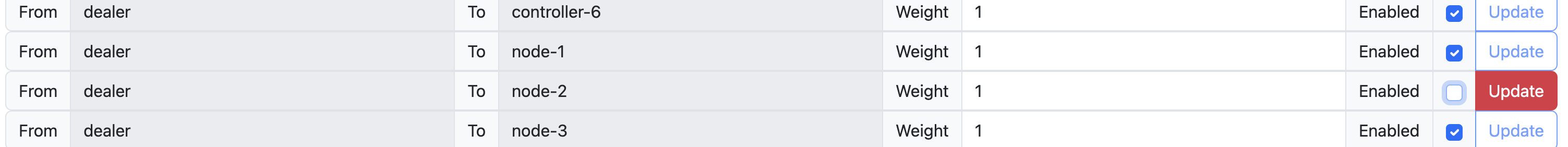}
        \caption{Disable a link between the dealer and a node.}\label{fig:DisableLink}
    \end{center}
 \end{figure}

    \item \textbf{Routes config} \\ 
    The routes configuration allows the user to edit the links or paths between the services. By default, all services (Controller, Nodes, and dealer) are connected to each other with a weight or distance of one, see Fig.~\ref{fig:RoutesConfigUI}. To simulate a broken link between two services, we can either increase the weight to a high value, or set the link as disabled. For example, if we disable the link between dealer to node-1, traffic may be routed to node-2, then from node-2 to node-1, see Fig. \ref{fig:DisableLink}.

\subsection{Use case 1: simulate broken path}

\begin{figure}
    \begin{center}
        \includegraphics[scale=0.25]{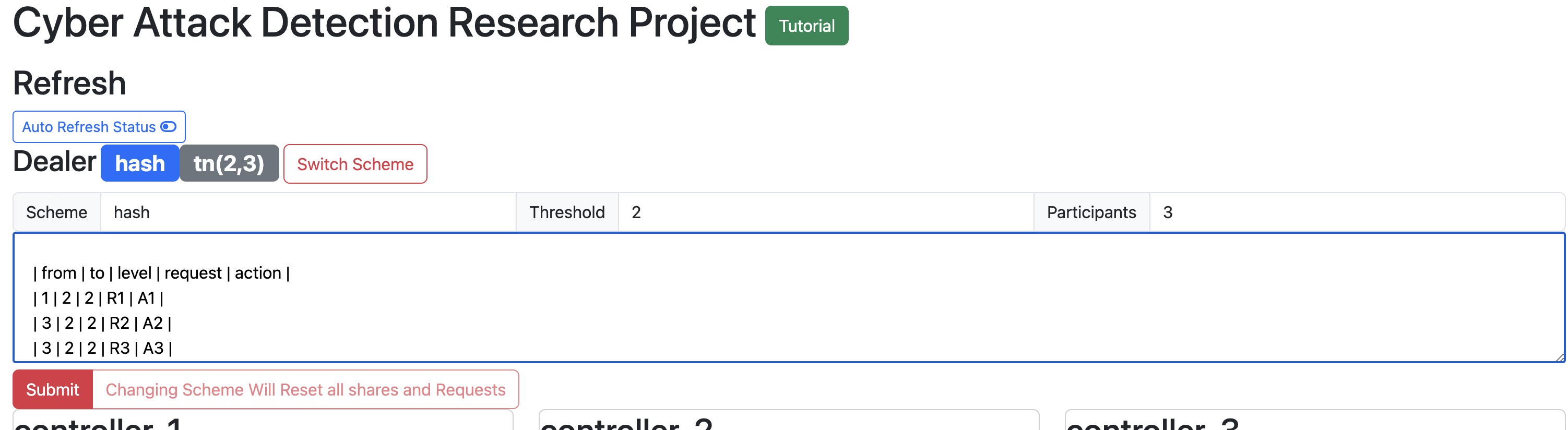}
        \caption{Use case: make choice of thrshold scheme.}\label{fig:Usecase_config}
    \end{center}
 \end{figure}

\begin{figure}
    \begin{center}
        \includegraphics[scale=0.25]{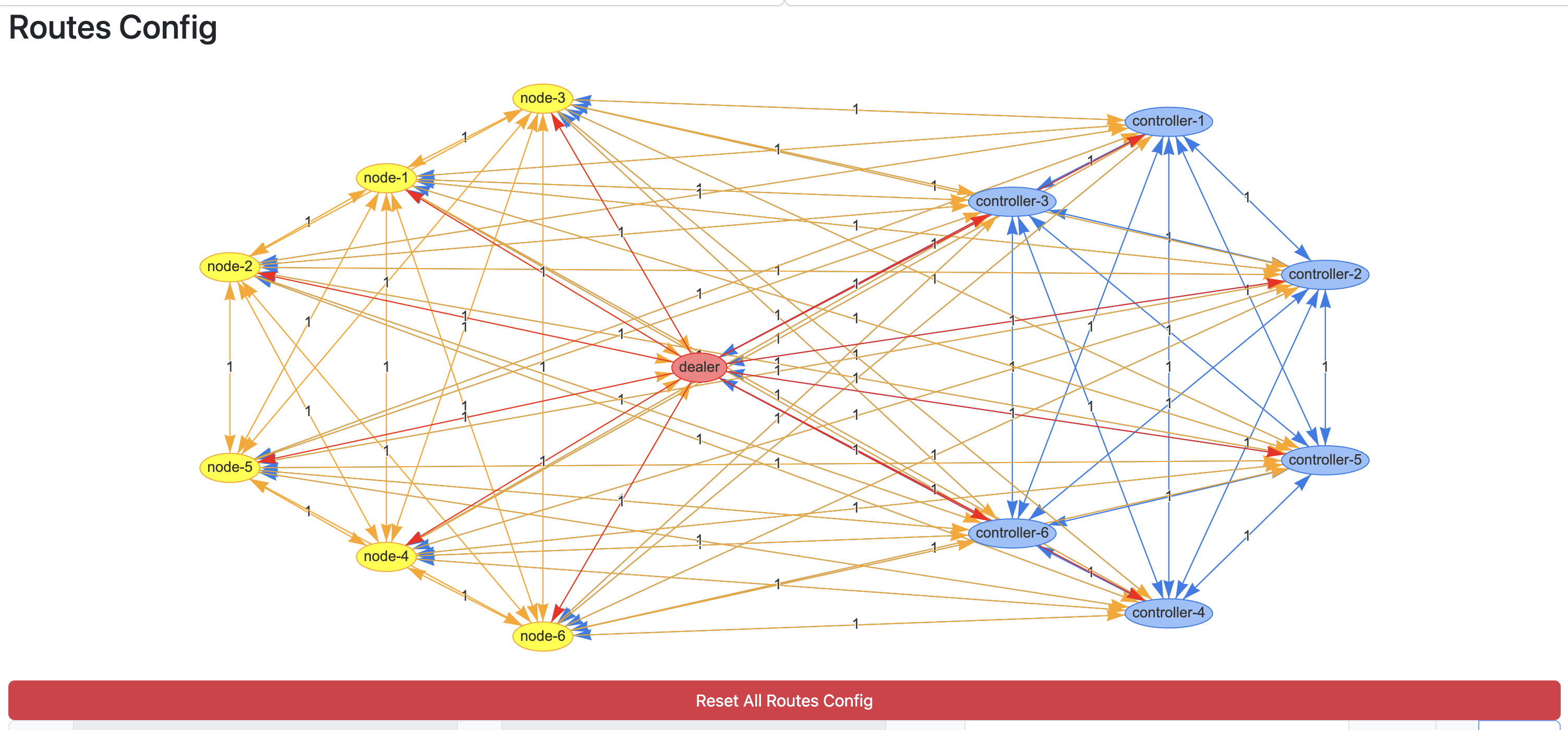}
        \caption{Use case: reset previous route configurations.}\label{fig:Usecase_resetrouteconfig}
    \end{center}
 \end{figure}

\begin{figure}
    \begin{center}
        \includegraphics[scale=0.35]{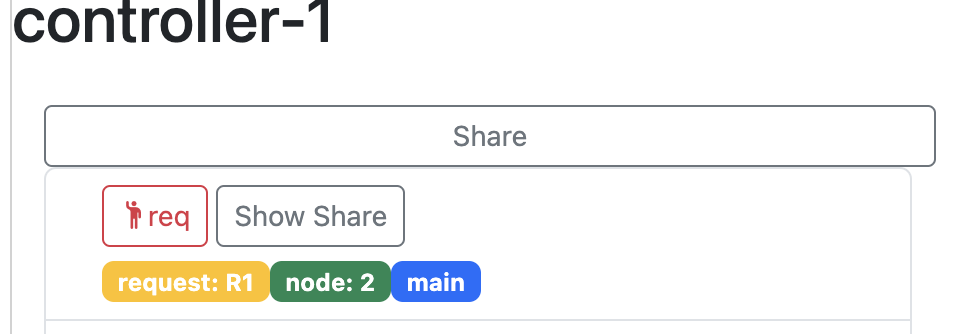}
        \caption{Use case: raise a request.}\label{fig:Usecase_raise_request}
    \end{center}
 \end{figure}

\begin{figure}
    \begin{center}
        \includegraphics[scale=0.25]{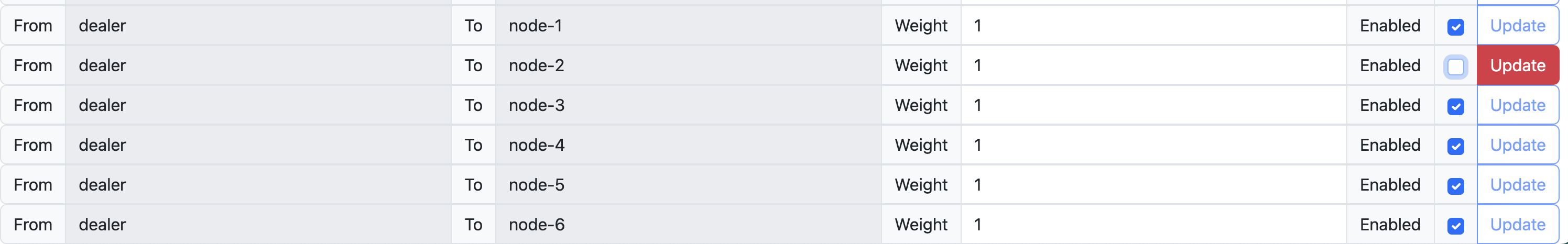}
        \caption{Use case: disable a path from route configuration table.}\label{fig:Usecase_disable_config_table}
    \end{center}
 \end{figure}

\begin{figure}
    \begin{center}
        \includegraphics[scale=0.25]{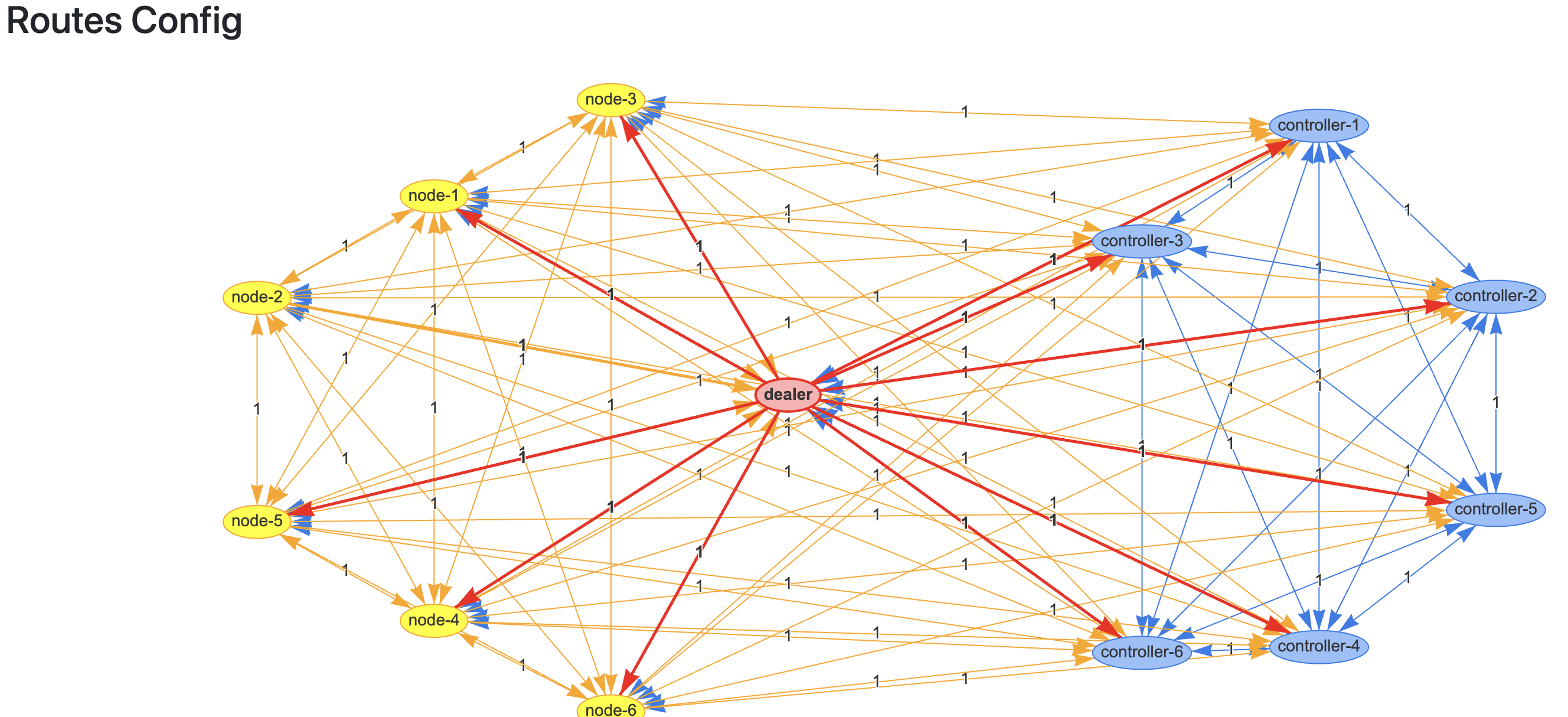}
        \caption{Use case: verify disabled path on the route graph.}\label{fig:Usecase_route_graph}
    \end{center}
 \end{figure}

\begin{figure}
    \begin{center}
        \includegraphics[scale=0.25]{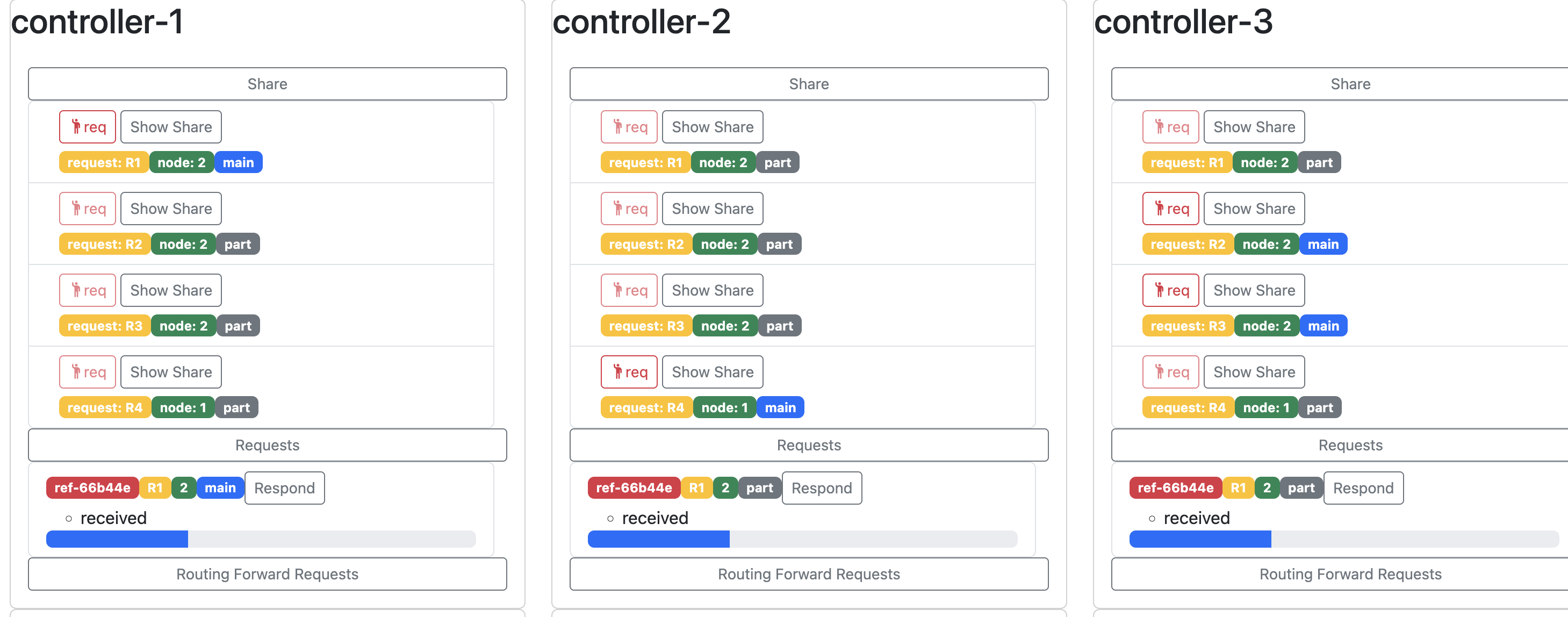}
        \includegraphics[scale=0.25]{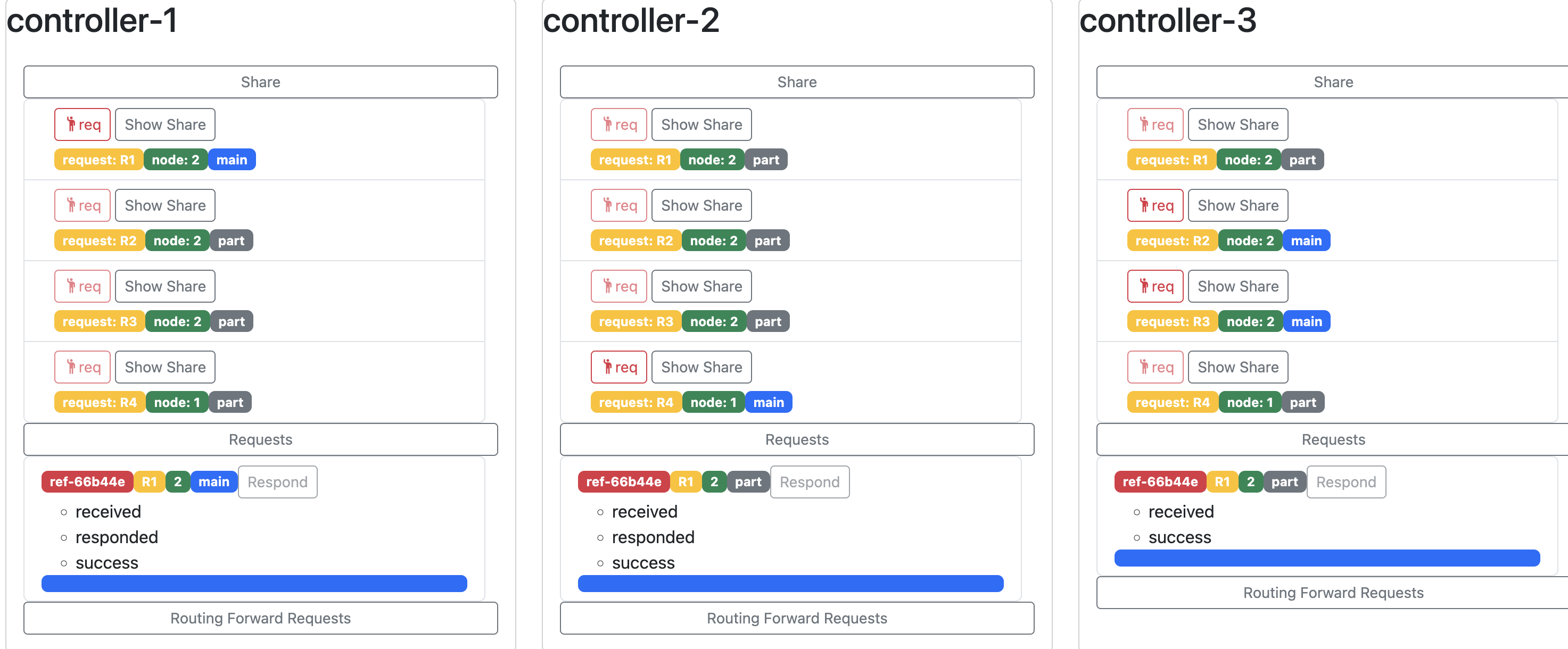}
        \caption{Use case: recover secret using two controllers.}\label{fig:Usecase_secret_recovery}
    \end{center}
 \end{figure}

\begin{figure}
    \begin{center}
        \includegraphics[scale=0.25]{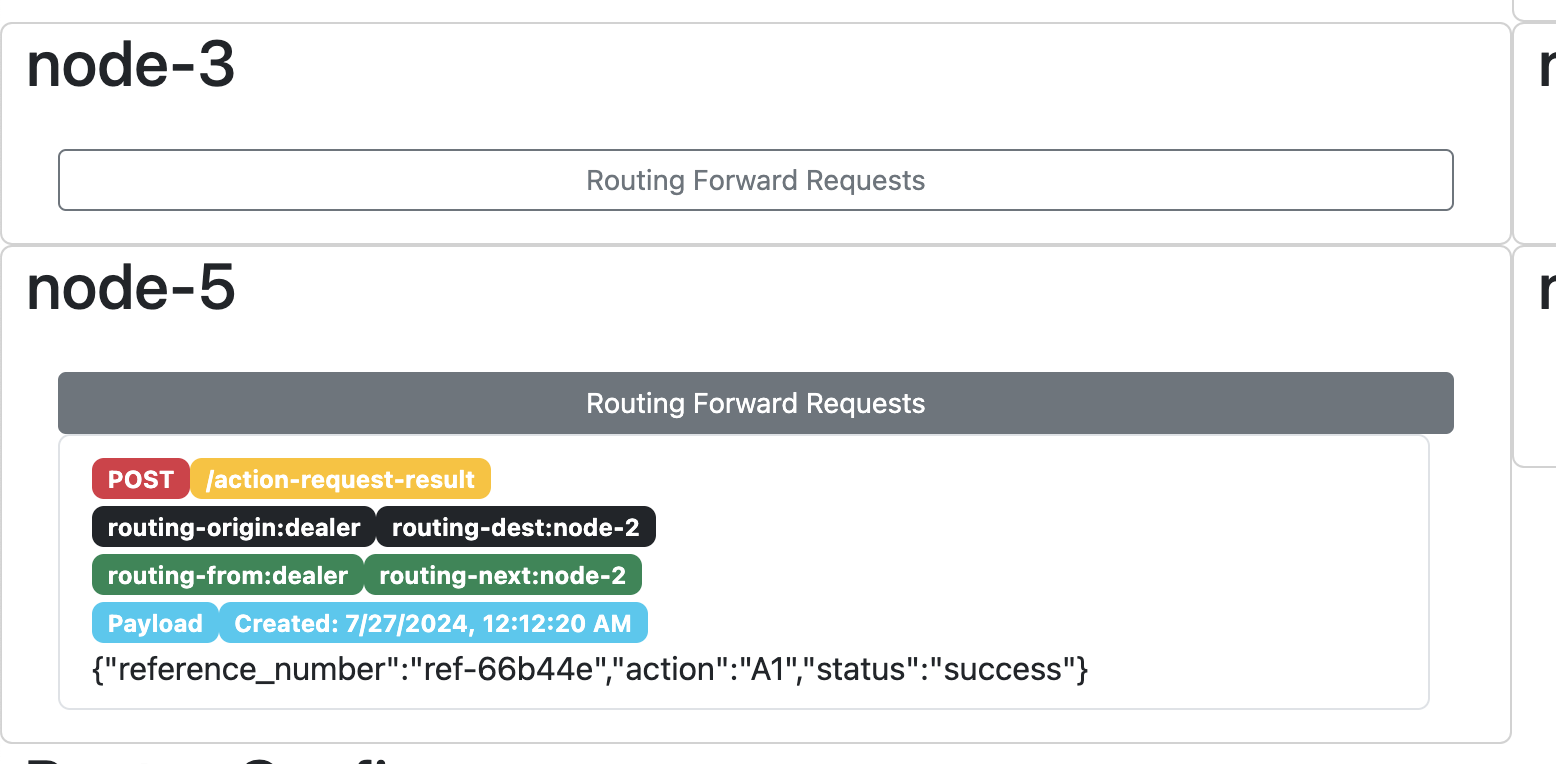}
        \caption{Use case: verify routing forward requests.}\label{fig:Usecase_forward_request}
    \end{center}
 \end{figure}

We will simulate a broken path between the dealer and a node. The end result will be that the node will receive the recovery result from another node, a so-called ``passthrough'' request, or forwarded request. We use Fig. \ref{fig:Usecase_config} through Fig. \ref{fig:Usecase_forward_request} to illustrate this use case. 


\begin{itemize}
\item Fig. \ref{fig:Usecase_config}: First we use Hash-based (2, 3)-threshold scheme, in which we only need two shares out of three participants to recover the secret. 
\item Fig. \ref{fig:Usecase_resetrouteconfig}: We also make sure we reset any previous route configurations.  
\item Fig. \ref{fig:Usecase_raise_request}: Next, we will click on Controller 1 to raise a request which the action node is Node-2. 
\item Fig. \ref{fig:Usecase_disable_config_table}: Then, we simulate a broken path between the dealer and Node-2 by disabling it in the routes configuration table.
\item Fig. \ref{fig:Usecase_route_graph}: After updating it, we can also refresh the page to verify that the route is disabled on the Routes Graph. 
\item Fig. \ref{fig:Usecase_secret_recovery}: We continue as normal to let the controllers respond to the request. We only need two controllers for this request to recover.  
\item Fig. \ref{fig:Usecase_forward_request}: Finally, we can see Node-2 actually performs an action. Since we disabled the path from dealer to Node-2, the recovery result should have traveled through some other node before reaching Node-2. We can see that Node-5's Routing Forward Requests actually has one entry with the same reference number. That it was originated from dealer, and target to Node-2.  
\end{itemize}

\section{Conclusions and future work}\label{Conclusions}

\begin{enumerate}[$\cdot$] 

\item We apply secret sharing schemes, especially hash function based schemes, into a real-world cyberattack detection model. The model runs an algorithm with multiple components and a database, which stores records of requests and their allowed actions.    

\item The model implementation specifies the interconnections of the components and allowed request types. In the setup phase we consider no-restrictions and with-restrictions (level-1 and level-2) cases between a controller and node.

\item The model implementation is split into the frontend UI, backend individual services, and a persistent layer of a single Postgres Database. The UI serves as the end-user dashboard page and direction calls from dashboard to corresponding backend services. Backend, written in Python, consists of four main services, namely, dealer, controller, node, and name registry. A free database PostgreSQL is used to store actions and logs for recovery attempts. For running environment, Docker is used to manage the start and stop of individual services and their URLs.

\item The model is implemented on Google Cloud Platform with graphic user interface. We demonstrate how to use the UI to simulate a complete flow of action request and secret sharing schemes. We explain UI components and their functions, dealer service, controllers, and nodes. We also give a use case of simulation of broken path in details.

\item The project demo is hosted on Computer Science Department's website permanently. We will use it as an example to promote and enhance student research involvement in cybersecurity and related areas for all levels of students from high schoolers to graduate students. We'll also integrate our research activities into the teaching to motivate more students to do research. It can be used as a few course modules for our undergraduate and graduate cybersecurity, network security, cryptography, cloud computing, and/or software engineering courses.  

\item Future work includes identifying compromised controllers based on a series of audit histories, or before we reissue new shares, we would ask each controller to provide their existing shares and validate their authenticity.  
\end{enumerate}

\section*{Funding}

This work was supported in part by a grant (Jul 2023 - Aug 2024) from Google Cyber NYC Institutional Research Program.


\bibliographystyle{plain}


\end{document}